\documentclass[prl,aps,10pt,twocolumn,longbibliography,floatfix,groupedaddress,superscriptaddress]{revtex4-2}
\usepackage[normalem]{ulem}
\usepackage[dvipsnames]{xcolor}
\usepackage{graphicx}
\usepackage{verbatim}
\usepackage{bbm}
\usepackage{bm}
\usepackage{amsmath} 
\usepackage{amssymb}
\usepackage{mathptmx}
\usepackage[version=3]{mhchem}

\usepackage{dsfont}
\usepackage[colorlinks=true,linkcolor=Blue,citecolor=Blue,urlcolor=Blue]{hyperref}

\newcommand{\ba}{\begin{eqnarray}}
\newcommand{\ea}{\end{eqnarray}}
\newcommand{\bd}{\begin{displaymath}}

\newcommand{\nn}{\nonumber \\}

\newcommand{\Z}{{\mathbb Z}}

\newcommand{\SMsection}[2]{%
\refstepcounter{section}\setcounter{subsection}{0}%
\section*{\thesection. #1}\label{#2}}

\newcommand{\SMsubsection}[2]{%
\refstepcounter{subsection}\setcounter{subsubsection}{0}%
\subsection*{\thesubsection. #1}\label{#2}}

\newcommand{\SMsubsubsection}[2]{%
\refstepcounter{subsubsection}%
\subsubsection*{\thesubsubsection. #1}\label{#2}}

\begin{document}
\title{\texorpdfstring{Coexisting orders and dipolar Luttinger liquids in $\Z_N$ quantum torus chain}{Coexisting orders and dipolar Luttinger liquids in $\Z_N$ quantum torus chain}}

\author{Seunghun \surname{Lee}}
\thanks{These authors contributed equally to this work.}
\affiliation{Department of Physics, Korea Advanced Institute of Science and Technology, Daejeon 34141, South Korea}

\author{Seungho \surname{Lee}}
\thanks{These authors contributed equally to this work.}
\affiliation{Department of Physics, Sungkyunkwan University, Suwon 16419, South Korea}
\affiliation{Institute of Basic Science, Sungkyunkwan University, Suwon 16419, South Korea}

\author{Hyun-Yong \surname{Lee}}
\email[Co-corresponding author: ]{hyunyong@korea.ac.kr}
\affiliation{Division of Semiconductor Physics, Korea University, Sejong 30019, Korea}
\affiliation{Department of Applied Physics, Graduate School, Korea University, Sejong 30019, Korea}

\author{Jung Hoon \surname{Han}}
\email[Co-corresponding author: ]{hanjemme@gmail.com}
\affiliation{Department of Physics, Sungkyunkwan University, Suwon 16419, South Korea}
\affiliation{Center for Quantum Dynamics of Angular Momentum, Pohang University of Science and Technology, Pohang 37673, South Korea}

\begin{abstract}
We establish a unified ground-state phase diagram for the one-dimensional $\mathbb{Z}_N$ quantum torus chain, which is connected to dipolar cluster and dipolar Ising chains via noninvertible dualities. Combining tensor-network calculations with bosonization and perturbation theory, we establish a rich phase diagram governed by the local Hilbert-space dimension $N$ and its parity. We identify phases with coexisting $X$ and $Z$ long-range orders ($C$ phase), and dipolar Luttinger liquids (dLL$_{\rm I}$ and dLL$_{\rm II}$) separating coexisting phases from single-order phases. Gapless dipolar excitations coexist with long-range order in dLLs. Both the $C$ and dLL phases appear at $N \geq 5$, but only the $C$ phases persist as $N\to \infty$. The stability of the dLL phase is understood by an emergent Luttinger liquid  theory, with Berezinskii-Kosterlitz-Thouless transitions out of it driven by competing umklapp and clock anisotropy perturbations. For odd $N \geq 5$, a third dLL phase (dLL$_{\rm III}$) emerges near $\theta = \pi$ due to the low-energy dynamics inside the exponentially degenerate ground-state manifold that maps onto an effective XXZ chain. Duality considerations suggest a possible gapless dipolar symmetry-protected topological phase and spontaneous dipole-symmetry breaking in dipolar cluster chains.
\end{abstract}
\date{\today}
\maketitle

\textit{Introduction.}---Exploring the phase diagrams of simple quantum chains has long provided a route to understanding order and criticality in one dimension. The transverse-field Ising chain offers a canonical example, in which tuning the field drives a transition between ordered and paramagnetic phases~\cite{pfeuty70the}. More intricate phase structures emerge when competing interactions are varied, as illustrated by the spin-1 bilinear--biquadratic chain~\cite{lauchli06nematics}. These examples highlight how even simple quantum chains can exhibit diverse forms of order and criticality as their competing interactions are tuned.

Another route to enriching the phase diagrams of quantum chains is to vary the local Hilbert-space dimension $N$ along with the interaction strengths. A paradigmatic example is the two-dimensional classical $\Z_N$ clock model and its one-dimensional quantum counterpart, where an intermediate critical phase emerges for $N \geq 5$~\cite{jose77renormalization,elitzur79phase,sun19phase}. A systematic investigation of order and criticality in quantum chains with both varying $N$ and competing interactions is largely lacking, with a few exceptions such as the quantum torus chain (QTC)~\cite{lee12quantum} featuring competing $X$ and $Z$ bond interactions and the transverse $\mathbb{Z}_N$ cluster chain~\cite{tsui17the}. Previous investigations have focused on small values of $N$, leaving its phase structure at larger $N$ unexplored.

A further motivation to revisit the QTC comes from the recent realization of its duality to the dipolar transverse Ising chain~\cite{cao24generating} and the dipolar transverse cluster model~\cite{han24topological,han25noninvertible}. Dipolar quantum systems have attracted considerable interest for their unusual dynamics and phases, including dipolar Luttinger liquids (dLLs) arising from constrained dipole motion in bosonic systems~\cite{zechmann23fractonic,lake23dipole}. These developments suggest that exploring the QTC at larger $N$ may provide insights into a broader range of dipolar quantum phenomena, which we indeed find to be the case.

\begin{figure}[t]
\centering
\includegraphics[width=\columnwidth]{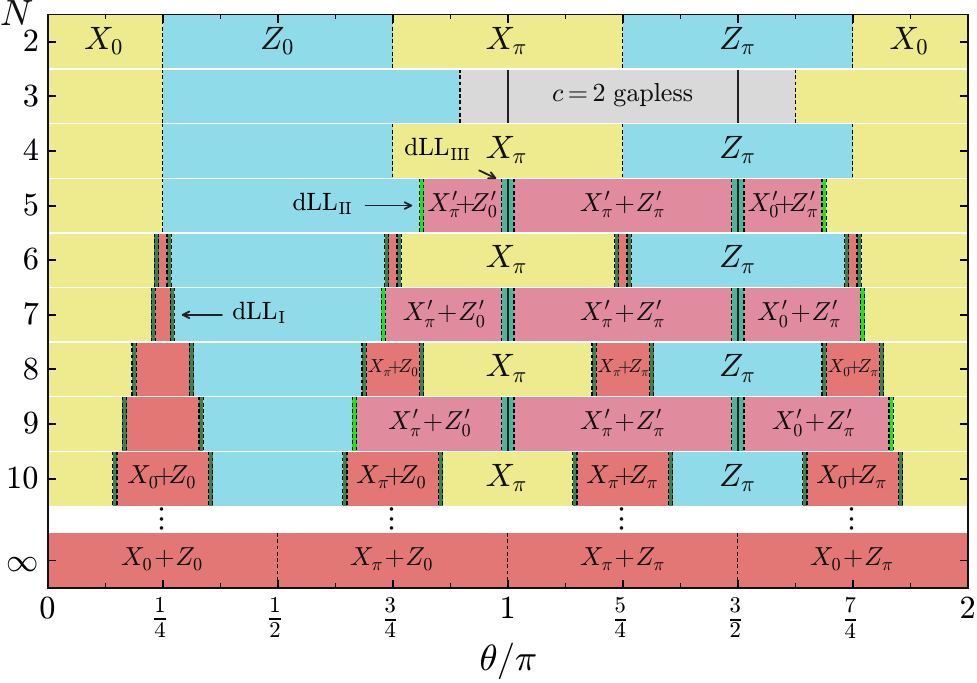}
\caption{Phase diagram of the $\Z_N$ quantum torus chain for $2\leq N\leq 10$ and its extrapolation to $N = \infty$. Yellow, blue, and red regions indicate $X_q$, $Z_q$, and $X_q + Z_{q'}$ phases, respectively, with $q, q' = 0, \pi$. Pink regions represent $X'_{q} + Z'_{q'}$. Narrow green regions indicate one of the three dipolar Luttinger liquids labeled dLL$_{\rm I}$, dLL$_{\rm II}$, and dLL$_{\rm III}$. For odd $N\geq 5$, dLL$_{\rm III}$ phases lie near the points $\theta = \pi$ and $3\pi/2$, where the ground states are exponentially degenerate and represented by black solid lines. The gray region for $N = 3$ denotes a gapless phase with central charge $c = 2$~\cite{lee12quantum}.}
\label{fig:phase-diagram}
\end{figure}

In this Letter, we show how increasing the local Hilbert-space dimension $N$ alone can re-balance the competition between $X$ and $Z$ orders in the QTC, revealing a rich phase structure governed by $N$ and its parity (see Fig.~\ref{fig:phase-diagram}). Combining tensor-network calculations with low-energy effective theories, we identify gapped phases with coexisting $X$ and $Z$ orders, as well as three types of dipolar Luttinger liquids (dLLs) where dipolar gapless fluctuations coexist with background long-range order. A dipolar-rotor theory explains the tendency toward coexistence and, together with bosonization, accounts for the stability of two types of dLLs. For odd $N\geq 5$, a third dLL emerges from a distinct mechanism involving constrained dynamics near an exponentially degenerate ground-state manifold. Through noninvertible dualities, we further explore the implications of these findings for the dual chains, suggesting the possibility of gapless dipolar SPT order and spontaneous dipole-symmetry breaking.

{\it Quantum torus chain.}---Consider a chain of $N$-dimensional local Hilbert spaces with basis states $\{ | g_j \rangle \}_{g_j \in \Z_N}$. The generalized Pauli operators $Z_j$ and $X_j$ act as $Z_j | g_j \rangle = \omega^{g_j} | g_j \rangle$ and $X_j | g_j \rangle = | g_j + 1 \rangle$, where $\omega = e^{2\pi i/N}$ and addition is modulo $N$. The QTC Hamiltonian ~\cite{lee12quantum}
\begin{equation} 
    H_{\rm QTC} (\theta) = -\sum_j \big( \cos\theta\, X_j X_{j+1}^\dagger + \sin\theta\, Z_j Z_{j+1}^\dagger + \mathrm{h.c.} \big)
    \label{eq:QTC}
\end{equation}
possesses two global $\Z_N$ unitary symmetries $U_X = \prod_j X_j$ and $U_Z = \prod_j Z_j$. The Hadamard duality exchanges $X_j$ and $Z_j$ (up to Hermitian conjugation) and implements $\theta \mapsto \pi/2 - \theta$, which allows us to restrict the phase diagram to, for instance, $\pi/4 \leq \theta \leq 5\pi/4$ for any $N$. For even $N$, additional unitary symmetries by $V_Z = \prod_{j, \mathrm{odd}} Z_j^{N/2}$ and $V_X = \prod_{j, \mathrm{odd}} X_j^{N/2}$ implement $\theta \mapsto \pi - \theta$ and $\theta \mapsto -\theta$, respectively, further reducing the range of independent parameters to $\pi/4 \leq \theta \leq \pi/2$. These transformations are absent for odd $N$, allowing a richer phase structure discussed below. 

{\it Overview of the phase diagram.}---We investigate the ground-state phase diagram of the QTC using infinite density-matrix renormalization group (iDMRG)~\cite{white92dmrg,mcculloch08infinite} and variational uniform matrix-product-state (VUMPS)~\cite{zaunderstauber18variational} calculations for $0 \le \theta < 2\pi$ and $2 \le N \le 10$. For a two-site unit cell, we define the order parameters
\begin{align}
    m_{O, q} = \frac 12 \left| \sum_{j=1}^2 e^{iqj} \langle O_j \rangle \right|, \quad O = X,Z .
\end{align} 
Only the uniform ($q = 0$) and staggered ($q = \pi$) orders were found, even for a larger unit cell. Phases with a single nonzero order are dubbed single-order (SO) phases and denoted $X_q$ or $Z_q$. Those with two or more orders are termed coexistence (C) phases and denoted $X_{q_1} + Z_{q_2}$ or $X'_{q_1} + Z'_{q_2}$. In the latter, the two dominant orders $X_{q_1}$ and $Z_{q_2}$ are accompanied by subdominant components $X_{q_1 +\pi}$ and $Z_{q_2 + \pi}$. 

The overall phase diagram is shown in Fig.~\ref{fig:phase-diagram}. Previous work identified several SO phases for $N = 2$ and an additional gapless phase with central charge $c = 2$ for $N = 3$~\cite{lee12quantum}. Extending the analysis to larger $N$, we uncover a richer phase structure featuring C phases and three types of dLLs, and a sensitive dependence on the parity of $N$.

In detail, the $X_0$ SO phase near $\theta = 0$ gives way to $Z_0$ SO phase past the self-dual point $\theta = \pi/4$ through a continuous ($N = 2$) or first-order ($N = 3, 4, 5$) transition. For $N \geq 6$, this direct transition is supplanted by an intervening $X_0 + Z_0$ C phase, whose width increases with $N$ to occupy the entire quadrant $0 < \theta < \pi/2$ as $N\to \infty$. Upon closer inspection, however, we discover that the $X_0 + Z_0$ phase is surrounded by thin slices of dLL$_{\rm I}$ phases on both sides. The sequence of transitions upon increasing $\theta$ from 0 to $\pi/2$ is therefore
\begin{align} 
    X_0 \to {\rm dLL}_{\rm I} \to X_0 + Z_0 \to {\rm dLL}_{\rm I} \to Z_0 . \nonumber
\end{align} 
For even $N$, this transition pattern is repeated under the aforementioned unitary transformations.

For odd $N \geq 5$, the C phases near $\theta = \pi/4$ are of the $X_{q_1} + Z_{q_2}$ type but of the $X'_{q_1} + Z'_{q_2}$ type elsewhere. The residual components in $X'_{q_1} + Z'_{q_2}$ originate from the frustration of antiferromagnetic clock interaction specific to odd $N$. In $\pi/2 < \theta < \pi$, a second dLL phase (dLL$_{\rm II}$) separates the $Z_0$ SO phase from the $X'_\pi + Z'_0$ C phase. A third dLL phase (dLL$_{\rm III}$) appears on both sides of $\theta = \pi$ where the ground states are exponentially degenerate. Another dLL$_{\rm III}$ phase appears near $\theta = 3\pi/2$ by duality. 

In all, dLL$_{\rm I}$ appears on both sides of $\theta = \pi/4$ for all $N \geq 6$, while dLL$_{\rm II}$ and dLL$_{\rm III}$ occur only for odd $N \geq 5$, with the former found in $\pi/2 < \theta < \pi$ and the latter near $\theta = \pi, 3\pi/2$. All dLL boundaries host Berezinskii-Kosterlitz-Thouless (BKT) transitions. In the $N\to \infty$ limit, our effective theory predicts that only the four $X_q + Z_{q'}$ C phases survive, each occupying one quadrant of the phase diagram.

{\it Dipolar-rotor theory of C phase.}---To understand how competing $X$ and $Z$ orders can coexist, we develop a large-$N$ rotor description around the $Z_0$ order at $\theta = \pi/2$. In the $Z$ eigenbasis $| m_j \rangle$ where $Z_j | m_j \rangle = e^{i\epsilon m_j} | m_j \rangle$ with $\epsilon = 2\pi / N$, we relax the $\mathbb{Z}_N$ clock variables to become integers $m_j \in \Z$, and write the shift operator $X_j = e^{i\phi_j}$ with $[\phi_j, m_k] = i\delta_{jk}$. Expanding the $Z$-bond interaction of Eq.~\eqref{eq:QTC} for smooth $m$ configurations gives the dipolar quantum rotor (dQR) Hamiltonian
\begin{equation}
    H_{\rm dQR} = \sum_j (m_{j+1} - m_j)^2 - 2\kappa \sum_j \cos(\phi_{j+1} - \phi_j),
    \label{eq:dqr}
\end{equation}
with $\kappa = \cot\theta / \epsilon^2$. Its dipolar character follows from the action of $X_j X_{j+1}^\dagger$ in the $Z$-basis, which locally updates the bond variables $n_j = m_{j+1} - m_j$ by 
$(\delta n_{j-1}, \delta n_j, \delta n_{j+1}) = (1, -2, 1)$,
conserving both the total charge $\sum_j n_j$ and dipole moment $\sum_j j n_j$~\footnote{On a periodic chain of length $L$, a hop across the boundary changes the integer dipole moment $D = \sum_j j n_j$ by $\pm L$. Thus, $D$ is conserved modulo $L$ for integer rotors, or modulo $\gcd (N, L)$ for $\Z_N$ clock variables. Full $\Z_N$ dipole conservation is compatible with periodic boundaries when $N$ divides $L$.}.

\begin{figure}[t]
    \includegraphics[width=\columnwidth]{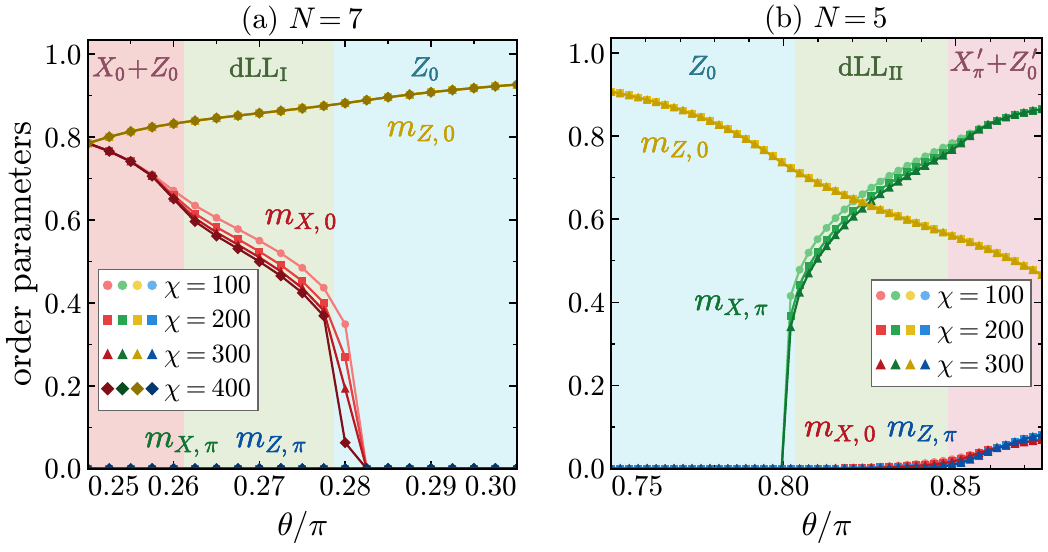}
    \caption{Uniform and staggered order parameters for (a) $N = 7$ near dLL$_{\rm I}$ and (b) $N = 5$ near dLL$_{\rm II}$. Red, green, yellow, and blue symbols represent $m_{X,0}$, $m_{X,\pi}$, $m_{Z,0}$, and $m_{Z,\pi}$ order, respectively. Circles, squares, triangles, diamonds correspond to bond dimensions $\chi = 100, 200, 300, 400$. Shaded regions indicate $Z_0$ (blue), $X_0 + Z_0$ (red), $X'_{\pi} + Z'_0$ (pink), and dLL (green) phases. Within the dLL regions, $Z_0$ order persists, whereas the apparent (a) $X_0$ and (b) $X_\pi$ orders at finite $\chi$ are expected to vanish as $\chi \to \infty$.}
    \label{fig:op_curve}
\end{figure}

Expanding Eq.~\eqref{eq:dqr} around the saddle points $\phi_j = qj + \varphi_j$, with $q = 0$ ($\kappa > 0$) or $\pi$ ($\kappa < 0$), yields the harmonic Hamiltonian with asymptotic correlations $\langle Z_j Z_{j+r}^\dagger \rangle$ and $e^{iqr} \langle X_j X_{j+r}^\dagger \rangle$ approaching $e^{-\epsilon^2 \sqrt{|\kappa|}/2}$ and $e^{-1/{2 \sqrt{|\kappa|}}}$ as $r\to \infty$~\cite{SM}. At $\theta$ away from $\pi/2$, both exponents scale as $1/N$ and the two correlators approach a finite value, explaining the growing tendency toward coexistence with increasing $N$ shown in Fig.~\ref{fig:phase-diagram}. The simultaneous nonzero values of $m_{X,0}$ and $m_{Z,0}$ in the $X_0 + Z_0$ C phase are numerically confirmed in Fig.~\ref{fig:op_curve}(a).

For fixed $N$, the exponent in the $X_q$ correlator grows as $\theta$ approaches $\pi/2$,  indicating strong fluctuation in the phase $\phi_j$ and the eventual loss of $X_q$ order at $\kappa < \kappa_c$. The corresponding critical angle is determined by $\kappa_c = (N/2\pi)^2 \cot\theta_c$, giving  $\theta_c \simeq \pi/2 - 4\pi^2 \kappa_c / N^2$. By $X \leftrightarrow Z$ duality, the corresponding boundary on the opposite side is $\pi/2 - \theta_c$. Thus, the intervening $C$ phase widens with increasing $N$, as observed in Fig.~\ref{fig:phase-diagram}. A Pad\'e analysis gives $\kappa_c \simeq 1.15$~\cite{SM}, which fits the C to SO phase boundaries well at large $N$. 

\begin{figure}[t]
    \includegraphics[width=\columnwidth]{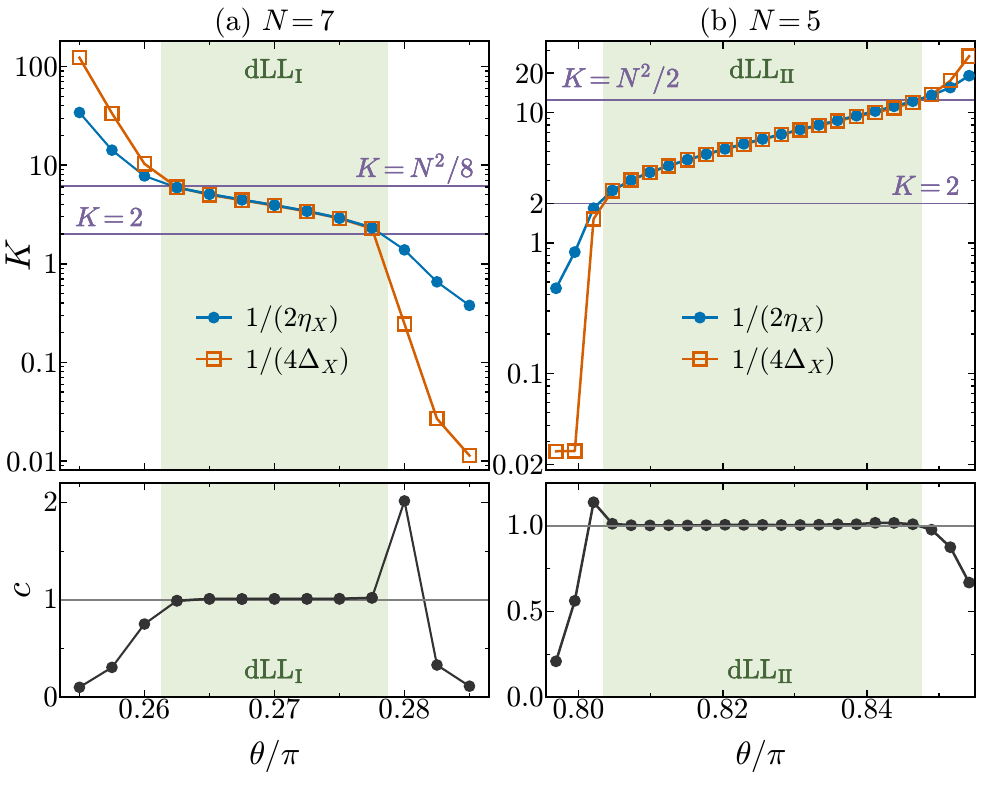}
    \caption{Numerical evidence for (a) dLL$_{\rm I}$ ($N = 7$) and (b) dLL$_{\rm II}$ ($N = 5$). Upper panels: Luttinger parameter $K$ extracted from the power-law fits of $\langle X_j X_{j+r}^\dagger \rangle$ (blue circles) and finite-entanglement scaling of the $X_q$ pseudo-order (orange squares), yielding $K = 1 / (2\eta_X)$ and $K = 1 / (4\Delta_X)$, respectively. Purple lines mark the stability windows of Eq.~\eqref{eq:K_window}. Lower panels: Central charge $c$ from the entanglement-entropy scaling $S_\chi \sim (c/6) \log \xi_\chi$. The green-shaded dLL regions exhibit consistent $K$ estimates and $c\simeq 1$. Correlation fits use $r\in[10,100]$ for both $N=7$ and $N=5$. Finite-entanglement scaling uses $\chi\in\{100,200,300,400\}$ for $N=7$ and $\chi\in\{100,200,300\}$ for $N=5$.}
    \label{fig:dLL_evidence}
\end{figure}

Numerically, the entanglement entropy (EE) saturates with increasing bond dimension $\chi$ in the SO and C phases~\cite{SM}, indicating their gapped nature. A narrow gapless region intervenes between the two phases (green strip in Fig.~\ref{fig:phase-diagram}), in which the $Z_0$ order persists. The $X_q$ order, on the other hand, decreases and is expected to ultimately vanish as the bond dimension $\chi$ increases -- see Fig.~\ref{fig:op_curve} and \cite{SM} for thorough numerical data to support the claim. Moreover, the scaling of EE with $\chi$ in the claimed critical regions yields a central charge $c\simeq 1$ (see Fig.~\ref{fig:dLL_evidence}). We identify the critical phases as dLL$_{\rm I}$ ($\theta < \pi/2$) and dLL$_{\rm II}$ ($\theta > \pi/2$). The harmonic approximation to Eq.~\eqref{eq:dqr} gives dynamical exponent $z = 2$ in contrast to $z=1$ of the dLL phase. We now develop an effective theory for dLL and its BKT transition into gapped phases. 

{\it Dipolar Luttinger liquids and their stability.}---We introduce continuum fields $\rho (x)$ and $\Theta (x)$ for the coarse-grained rotor number $m_j$ and phase fluctuation $\varphi_j$, respectively. Defining the field $\Phi (x)$ through $\rho(x) = \rho_0 + \partial_x \Phi / \pi$, where $\rho_0 = \langle m_j \rangle$ is the uniform background, and the discreteness of the microscopic $m_j$ allows the commensurability potential $V(\rho) = -y \cos(2\pi\rho)$, for $y > 0$. Choosing one of the integer background $\rho_0 \in \Z$ and expanding around it gives a gradient correction to the harmonic continuum theory of dQR as~\cite{SM}
\begin{align}
    H_{\rm eff} = \int dx \left[ \frac{1}{\pi^2} (\partial_x^2 \Phi)^2 + |\kappa| (\partial_x \Theta)^2  + 2g (\partial_x \Phi)^2 \right] . 
    \label{eq:perturbed_Lifshitz}
\end{align}
The $(\partial_x \Phi)^2$ term dominates over the higher-derivative term $(\partial_x^2 \Phi)^2$ at long wavelengths, yielding a linear dispersion $\omega(k) \sim v|k|$ with $v = \pi \sqrt{8 |\kappa| g}$. The resulting low-energy theory describes the Luttinger liquid with a Luttinger parameter $K = \sqrt{|\kappa| / (2g)}$~\cite{SM}. The low-energy representation $X_j \sim e^{iqj} e^{i\Theta}$ yields algebraically decaying correlations $\langle X_j X_{j+r}^\dagger \rangle \sim  e^{-iqr} r^{-\eta_X}$ with $\eta_X = 1/(2K)$. Since $X_j$ creates a local dipole of opposite bond charges, $(\delta n_{j-1}, \delta n_j) = (1, -1)$, the algebraic correlations represent gapless dipolar fluctuations and gives the name dipolar Luttinger liquid. 

The stability of the dLL is governed by two competing locking perturbations arising from umklapp processes and clock anisotropy,
\begin{equation}
    \delta H = \int dx [ \lambda_u \cos (2a\Phi) + \lambda_c \cos (b\Theta) ],
    \label{eq:locking_perturbations}
\end{equation}
where $a, b\in \Z$ are constrained by microscopic symmetries. For the integer-commensurate $Z_0$ background ($\rho_0 \in \Z$), translation symmetry allows the leading umklapp harmonic $a = 1$. The global $\Z_N$ symmetry requires $b \equiv 0 \pmod N$. For uniform order ($q = 0$), the leading anisotropy has $b = N$. For staggered order ($q = \pi$), translation symmetry further requires even $b$, giving $b = 2N$. The scaling dimensions are $\Delta_u = a^2 K$ and $\Delta_c = b^2 / (4K)$, respectively. Therefore, both perturbations are irrelevant when
\begin{equation}
    \frac{2}{a^2} < K < \frac{b^2}{8}, \quad (ab >4) 
    \label{eq:K_window}
\end{equation}
stabilizing the dLL within this compact-boson description.

Outside this window, the relevant locking term generally opens a gap by pinning the corresponding field. For $K < 2 / a^2$, umklapp pins the $\Phi$ field, disordering its conjugate field $\Theta$ and rendering $X_q$ correlations short-ranged. This is consistent with the gapped $Z_0$ SO phase. For $K > b^2 / 8$, clock anisotropy pins the $\Theta$ field and establishes long-range $X_q$ order on the existing $Z_0$ background. Thus, the dLL separates two distinct ordered regimes with a gap. Unlike dipole-conserving Bose--Hubbard models~\cite{lake22dipolar,lake23dipole,zechmann23fractonic}, the QTC possesses only discrete $\Z_N$ symmetries, yet supports a stable dLL.

We numerically verify the dLL behavior through finite-entanglement scaling and correlation functions. Within the dLL regions of Fig.~\ref{fig:op_curve}, the apparent $X_q$ order is a finite-entanglement pseudo-order that vanishes as $m_{X,q} \sim \xi_\chi^{-\Delta_X}$, where $\xi_\chi$ is the MPS correlation length at bond dimension $\chi$~\cite{SM}. Independently, the $XX$ correlations decay as $\langle X_j X_{j+r}^\dagger \rangle \sim e^{-iqr} r^{-\eta_X}$. The LL relation $2\Delta_X = \eta_X = 1 / (2K)$ yields consistent estimates of the Luttinger parameter $K$ within the predicted stability windows of Eq.~\eqref{eq:K_window}, as shown in Fig.~\ref{fig:dLL_evidence}. Furthermore, Fig.~\ref{fig:dLL_evidence} shows the central charge estimate from EE scaling $S_\chi \sim (c/6) \log \xi_\chi$, yielding $c \simeq 1$ within the same regions. Together, these results provide strong numerical evidence for the Luttinger-liquid nature of both dLL$_{\rm I}$ and dLL$_{\rm II}$. Further numerical evidence supporting the dLL picture is provided in the SM~\cite{SM}.

{\it Dipolar Luttinger liquid near $\theta = \pi$ and odd $N$.}---We now turn to dLL$_{\rm III}$ near $\theta = \pi$ for odd $N \geq 5$, whose microscopic origin is distinct from the dQR description above. For odd $N$ at $\theta = \pi$, each bond has two degenerate energy minima. In the $X$ eigenbasis, $X_j | s_j \rangle = e^{-i\epsilon s_j} | s_j \rangle$, the bond variables $l_j = s_{j+1} - s_j$ take the values $l_j = \pm N'$ with $N' = (N - 1) / 2$. These two choices per bond generate an exponentially degenerate ground-state (EDGS). A finite $\delta = \theta - \pi$ lifts this degeneracy through the $Z_j Z_{j+1}^\dagger$ interaction, which locally updates $(\delta l_{j-1}, \delta l_j, \delta l_{j+1}) = (1, -2, 1)$. First-order hopping can take place within the EDGS for $N=3$. For $N\geq 5$, it happens at second order in hopping through the $0110 \leftrightarrow 1001$ process where $| 0 \rangle_j \equiv | {+} N' \rangle_j$ and $| 1 \rangle_j \equiv | {-} N' \rangle_j$ are the two pseudospin degrees of freedom. With pseudospin operators such as $\sigma_j^z  | 0/1 \rangle_j = \pm | 0/1 \rangle_j$, the effective Hamiltonian inside EDGS is 
\begin{equation}
    \begin{aligned}
        H_{\rm eff} &= -\sum_j \Big[ J_{1,N} \sigma_j^z \sigma_{j+1}^z + J_{2,N} \sigma_j^z \sigma_{j+2}^z \\
        &\hspace{60pt} + \sigma_{j-1}^{+} \sigma_j^{-} \sigma_{j+1}^{-} \sigma_{j+2}^{+} + \mathrm{h.c.} \Big],
    \end{aligned}
    \label{eq:2nd_order_eff_Ham}
\end{equation}
where $\sigma_j^\pm = (\sigma_j^x \pm i \sigma_j^y) / 2$ and the coefficients $J_{1,N}$ and $J_{2,N}$ are given in the SM~\cite{SM}. This Hamiltonian conserves the total pseudospin charge $\sum_j \sigma^z_j$ and dipole moment $\sum_j j \sigma^z_j$, and displays fragmentation of the Hilbert space into disconnected Krylov sectors~\cite{sala20fragmentation,lydzba24local}.

Among the Krylov sectors of $H_{\rm eff}$ we identify a mobile sector in which each pair of adjacent bond pseudospins $(2j, 2j+1)$ is either $01$ or $10$. Representing each pair by a block spin, $| {\uparrow} \rangle_{j'} = | 01 \rangle_{2j, 2j+1}$ and $| {\downarrow} \rangle_{j'} = | 10 \rangle_{2j,2j+1}$ where $j' \equiv (2j, 2j+1)$, the process $01|10 \leftrightarrow 10|01$ becomes a nearest-neighbor block-spin exchange, reducing Eq.~\eqref{eq:2nd_order_eff_Ham} to the XXZ Hamiltonian for $N\ge 5$~\cite{SM}:
\begin{equation}
    \begin{gathered}
        H_{\rm XXZ} \propto \sum_{j'} \left( S_{j'}^x S_{j'+1}^x + S_{j'}^y S_{j'+1}^y + \Delta_N S_{j'}^z S_{j'+1}^z \right), \\
        \Delta_N = -\frac{\alpha_N (\alpha_N + 3)}{(\alpha_N + 1)(\alpha_N + 2)}, \quad \alpha_N = 1 + 2 \cos \frac{2\pi}{N}. 
    \end{gathered}
    \label{eq:mobile_sector_XXZ}
\end{equation}
Since $|\Delta_N| < 1$ for $N\geq 5$, the ground state of Eq.~\eqref{eq:mobile_sector_XXZ} is a $c = 1$ Luttinger liquid with~\cite{giamarchi04quantum,haldane81luttinger}
\begin{align}
    K_N = \frac{\pi}{2(\pi - \arccos \Delta_N)},
    \label{eq:Luttinger_K_2nd}
\end{align}
The nature of the gapless excitations is $0110\leftrightarrow 1001$ in the original pseudospin picture, which is dipole-conserving and qualifies the phase as dLL$_{\rm III}$. 

\begin{figure}[tbp]
\centering
\includegraphics[width=\columnwidth]{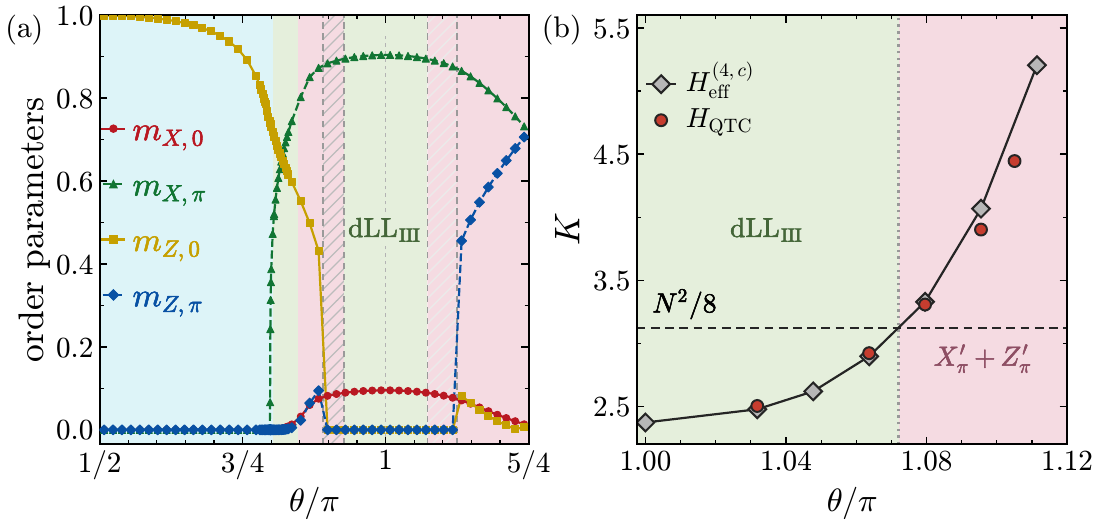}
\caption{(a) Uniform and staggered order parameters for $N = 5$ over $1/2 \leq \theta / \pi \leq 5/4$. Hatched regions indicate uncertainties in locating the phase boundary at finite $\chi$ and are expected to narrow as $\chi$ increases. (b) Luttinger parameter $K$ for $\theta > \pi$, obtained from the fourth-order effective Hamiltonian $H_{\rm eff}^{(4,{\rm c})}$ (gray diamonds) and the microscopic QTC (red circles), using the power-law fitting of $\langle Z_0 Z_r^\dagger \rangle$. The horizontal dashed line marks the clock-locking threshold $K = 25/8$, crossed near $\theta_{\mathrm{BKT}}^+ \simeq 1.073 \pi$ (vertical dotted line).}
\label{fig:dLL_near_pi}
\end{figure}

Figure~\ref{fig:dLL_near_pi}(a) shows coexisting $X_0$ and $X_\pi$ orders within the numerically identified dLL$_{\rm III}$ phase near $\theta = \pi$, with the $X_\pi$ component dominating. Since each block contains either $01$ or $10$, we have $l_{2j} + l_{2j+1} = 0$, implying $s_{2j} = f$ and $s_{2j+1} = f \pm N'$ with fixed $f\in \Z_N$. For the zero-magnetization XXZ ground state, the order parameters thus become $m_{X,0} = ( 1 - \cos \frac{\pi}{N} )/2, m_{X,\pi} = ( 1 + \cos \frac{\pi}{N} )/2$. For $N = 5$, these predict $m_{X,0} \simeq 0.0955$ and $m_{X,\pi} \simeq 0.9045$, in quantitative agreement with Fig.~\ref{fig:dLL_near_pi}(a). Thus, the second-order effective theory Eq.~\eqref{eq:mobile_sector_XXZ} accounts for both the dLL and its background orders.

To examine the stability of dLL$_{\rm III}$, we apply the compact-boson criterion in Eq.~\eqref{eq:K_window}. For the block-spin XXZ model, the zero-magnetization sector corresponds to half-filling, where translation symmetry forbids the leading umklapp $\cos (2\Phi)$ and instead allows $\cos (4\Phi)$, giving $a = 2$. Meanwhile, $Z_{2j+1} \sim e^{i\Theta}$ within the mobile sector, so the microscopic $\Z_N$ symmetry permits the leading clock anisotropy $\cos (N\Theta)$ with $b = N$~\cite{SM}. The resulting stability window is therefore $1/2 < K < N^2 / 8$.

At second order, $K_N$ [Eq.~\eqref{eq:Luttinger_K_2nd}] is independent of $\delta$. We therefore extend the perturbative analysis to fourth order for $N = 5$ and numerically determine $K_5 (\delta)$ from the two-point correlations of the fourth-order effective Hamiltonian $H_{\rm eff}^{(4,{\rm c})}$~\cite{SM}. As shown in Fig.~\ref{fig:dLL_near_pi}(b), the resulting $K_5 (\delta)$ matches well with the microscopic QTC results for $\delta > 0$ and reaches the clock-locking threshold $25/8$ near $\theta_{\rm BKT}^{+} \simeq 1.073 \pi$. At this threshold, $\cos (5\Theta)$ becomes marginal, consistent with a BKT transition to the gapped $X'_\pi + Z'_\pi$ phase. The effective theory $H_{\rm eff}^{(4,{\rm c})}$ also satisfies $K_5 (\delta) = K_5 (-\delta)$, suggesting a corresponding transition toward $X'_\pi + Z'_0$ for $\delta < 0$, although the full QTC has no exact reflection symmetry about $\theta = \pi$. In the SM, we provide an independent five-site $Z$-correlator diagnostic supporting the same BKT criterion~\cite{SM}. 

{\it Noninvertible duality and implications.}---The phase structure identified for the QTC has direct counterparts in models related by noninvertible dualities. In particular, the type-I dipolar transverse cluster-state model (dTCS) and the dipolar transverse Ising model (dTIM) are connected to the QTC through the duality web~\cite{han24topological,cao24generating,han25noninvertible}
\begin{equation}
    H_{\rm dTCS,I} \xrightarrow{\mathcal M_{\rm I}} H_{\rm dTIM} \xrightarrow{K_{\rm KW}} H_{\rm QTC},
\end{equation}
where $H \xrightarrow{\mathcal N} H'$ implies $\mathcal N H = H' \mathcal N$, and explicit expressions for noninvertible $\mathcal M_{\rm I}, K_{\rm KW}$ can be found in the SM~\cite{SM}. Both dual models take the form
\begin{align}
    H = -\sum_j (\cos\theta\, A_j + \sin\theta\, X_j + \mathrm{h.c.}),
\end{align}
where $A_j = Z_{j-1}^\dagger Z_j^2 Z_{j+1}^\dagger$ (dTIM) or $A_j = (Z_{j-1} Z_j^\dagger) X_j (Z_j^\dagger Z_{j+1})$ (dTCS). Both preserve the bulk charge and dipole symmetries $Q = \prod_j X_j$ and $D = \prod_j X_j^j$. Although the transformations are noninvertible, the bulk transitions in QTC carry over to the corresponding symmetry and boundary-twist sectors in the dual models~\cite{SM}. Under the duality, the two-point correlations $\langle X_{i-1}^\dagger X_{i+r-1} \rangle_{\rm QTC}$ and $\langle Z_{i-1} Z_{i+r-1}^\dagger \rangle_{\rm QTC}$ in QTC map to the cluster-string and paramagnetic-string correlators $\langle \prod_{j=i}^{i+r-1} A_j \rangle_{\rm dTCS}$ and $\langle \prod_{j=i}^{i+r-1} X_j^\dagger \rangle_{\rm dTCS}$ in the dTCS, respectively.

At $\theta = 0$, the dTCS is an exactly solvable dipolar SPT model~\cite{han24topological}. In the dLL$_{\rm I, II}$ phases, cluster-string order remains long-ranged while the paramagnetic string decays algebraically. 
The duality counterparts of the dLL$_{\rm I, II}$ phases imply dipolar SPT order surviving in the background of gapless bulk excitations, suggesting a possible realization of a gapless SPT phase~\cite{scaffidi17gapless, parker18topological}. In the $X_0 + Z_0$ $C$ phase, both cluster and paramagnetic strings exhibit long-range order. This coexistence suggests spontaneous dipole-symmetry breaking in the dTCS. Specifically, the two-point correlation of the dipole-charged operator $B_j = Z_{j-1} Z_j^\dagger$ maps to that of $W_j = X_j^\dagger Z_j$ in the QTC. A sufficient condition for $|\langle W_j \rangle| > 0$, and hence for dipole-symmetry breaking in the dual model, is $|\langle X_j \rangle|^2 + |\langle Z_j \rangle|^2 > 1$~\footnote{Let $x = |\langle X_j^\dagger \rangle|$ and $z = |\langle Z_j \rangle|$ in a symmetry-broken ground state. Applying the Cauchy--Schwarz inequality to $X_j^\dagger - \langle X_j^\dagger \rangle$ and $Z_j - \langle Z_j \rangle$ gives $|\langle W_j \rangle - \langle X_j^\dagger \rangle \langle Z_j \rangle| \leq \sqrt{(1 - x^2)(1 - z^2)}$. Consequently, $|\langle W_j \rangle| \geq xz - \sqrt{(1 - x^2)(1 - z^2)}$, whose right-hand side is positive precisely when $x^2 + z^2 > 1$.}. At $\theta = \pi/4$, the harmonic dQR theory yields $|\langle X_j \rangle|^2 + |\langle Z_j \rangle|^2 \simeq 2e^{-\pi/N} > 1$ in the QTC with large $N$, supporting spontaneous dipole-symmetry breaking in the dTCS. The $C$ phases and various dLL phases in the QTC for large $N$ thus have intriguing implications for the corresponding phase diagram in the dipolar SPT and dipolar Ising models, which merit further investigation.


{\it Acknowledgments.}---Tensor network calculations were performed using TeNPy~\cite{hauschild24tensor} and MPSKit~\cite{devos26mpskitjl,devos25tensorkitjl} libraries. JHH was supported by the National Research Foundation of Korea (NRF) grant funded by the Korea government(MSIT) (Grant No. 2023R1A2C1002644 and  No. RS-2024-00410027). H.-Y.L was supported by the Basic Science Research Program through the National Research Foundation of Korea funded by the Ministry of Science and ICT [Grant No. RS-2023-00220471, RS-2025-16064392]. SL was supported by the National Research Foundation of Korea (NRF) grant funded by the Ministry of Science and ICT (RS-2024-00451261). SL thanks Eun-Gook Moon for his guidance and support.

\bibliography{QTC}

\clearpage
\onecolumngrid
\begin{center}
{\large\bfseries Supplemental Material}\\[4pt]
{\bfseries ``Coexisting orders and dipolar Luttinger liquids in $\Z_N$ quantum torus chain''}
\end{center}
\hypersetup{linkcolor=black}

This Supplemental Material provides details of the analytical and numerical results supporting the main text. Section~\ref{app:dipolar-rotor-theory} develops the dipolar rotor description of coexisting orders, discusses its crossover to dipolar Luttinger liquids, and provides a Pad\'e estimate of the single-ordering boundary. Section~\ref{app:even-odd} derives the effective XXZ description of the odd-$N$ QTC near $\theta=\pi$ and analyzes the stability of the resulting dLL phase using bosonization. Section~\ref{app:web_of_dualities} establishes the duality mappings connecting the dipolar cluster, dipolar Ising, and quantum torus chains, as well as the corresponding mappings of correlation functions. Finally, Sec.~\ref{app:add_numerical_evidence} presents additional numerical results supporting the main text.

\tableofcontents
\vspace{1cm}

\twocolumngrid
\setcounter{section}{0}\setcounter{subsection}{0}
\setcounter{equation}{0}\setcounter{figure}{0}\setcounter{table}{0}
\renewcommand{\thesection}{\Roman{section}}
\renewcommand{\thesubsection}{\thesection.\arabic{subsection}}
\makeatletter\renewcommand{\p@subsection}{}\makeatother
\renewcommand{\theequation}{S\arabic{equation}}
\renewcommand{\thefigure}{S\arabic{figure}}
\renewcommand{\thetable}{S\arabic{table}}
\renewcommand{\theHsection}{supp.\arabic{section}}
\renewcommand{\theHsubsection}{supp.\arabic{section}.\arabic{subsection}}
\renewcommand{\theHequation}{supp.\arabic{equation}}
\renewcommand{\theHfigure}{supp.\arabic{figure}}
\renewcommand{\theHtable}{supp.\arabic{table}}

\SMsection{Dipolar rotor theory: Coexistence and crossover to dipolar Luttinger liquids}{app:dipolar-rotor-theory}

\SMsubsection{Coexistence of long-range orders}{app:coexist-long-range-order}

We derive the long-range correlations underlying the coexistence of $X$ and $Z$ orders in the harmonic dipolar rotor theory. Recall that $Z_j = e^{i\epsilon m_j}$ and $X_j = e^{i\phi_j}$, with $[\phi_j, m_l] = i\delta_{jl}$ and $\epsilon = 2\pi/N$. The dipolar-rotor Hamiltonian reads
\begin{align}
    H_{\rm dQR} = \sum_j (m_{j+1} - m_j)^2 -2\kappa \sum_j \cos(\phi_{j+1} - \phi_j),
\end{align}
with $\kappa = \cot\theta / \epsilon^2$. Expanding around the saddle point $\phi_j = qj + \varphi_j$, where $q = 0$ for $\kappa > 0$ and $q = \pi$ for $\kappa < 0$, gives the harmonic Hamiltonian
\begin{align}
    H_{\rm harm} =\sum_j \big[ (m_{j+1} - m_j)^2 + |\kappa| (\varphi_{j+1} - \varphi_j)^2 \big],
\end{align}
up to an additive constant.

Introducing the Fourier modes $m_j = L^{-1/2} \sum_k e^{ikj} m_k$ and $\varphi_j = L^{-1/2} \sum_k e^{ikj} \varphi_k$, we rewrite the harmonic Hamiltonian as
\begin{align}
    H_{\rm harm} = 4\sum_{k\neq 0} \sin^2 \frac k2 (m_k m_{-k} + |\kappa| \varphi_k \varphi_{-k}),
\end{align}
where the $k = 0$ modes have been omitted since they do not contribute to the field differences entering the correlations below. In the Gaussian ground state, we obtain
\begin{align}
    \langle m_k m_{-k} \rangle = \frac{\sqrt{|\kappa|}}{2}, \qquad \langle \varphi_k \varphi_{-k} \rangle = \frac{1}{2 \sqrt{|\kappa|}}.
\end{align}
The fluctuations of the field differences are then
\begin{equation} \label{eq:fluctuations}
    \begin{aligned}
        \langle (m_{j+r} - m_j)^2 \rangle &= \frac 1L \sum_{k\neq 0} |e^{ikr} - 1|^2 \langle m_k m_{-k} \rangle = \sqrt{|\kappa|}, \\
        \langle (\varphi_{j+r} - \varphi_j)^2 \rangle &= \frac 1L \sum_{k\neq 0} |e^{ikr} - 1|^2 \langle \varphi_k \varphi_{-k} \rangle = \frac{1}{\sqrt{|\kappa|}},
    \end{aligned}
\end{equation}
where we used $L^{-1}\sum_{k\neq0}|e^{ikr}-1|^2=2$ for $0<r<L$, as appropriate for the thermodynamic limit.

Using $Z_j = e^{i\epsilon m_j}$ and $X_j = e^{i(qj + \varphi_j)}$, the $X$ and $Z$ two-point correlations can be expressed as
\begin{equation}
    \begin{aligned}
        \langle Z_j Z_{j+r}^\dagger \rangle &= \left\langle e^{i\epsilon (m_j - m_{j+r})} \right\rangle, \\
        e^{iqr} \langle X_j X_{j+r}^\dagger \rangle &= \left\langle e^{i (\varphi_j - \varphi_{j+r})} \right\rangle.
    \end{aligned}
\end{equation}
Since the ground-state fluctuations are Gaussian about the saddle point, we use $\langle e^{iA} \rangle = e^{-\langle A^2 \rangle / 2}$ for the field differences considered here. Substituting Eq.~\eqref{eq:fluctuations}, we obtain
\begin{equation}
    \begin{aligned}
        \lim_{r\to \infty} \langle Z_j Z_{j+r}^\dagger \rangle &= e^{-\epsilon^2 \sqrt{|\kappa|} / 2}, \\
        \lim_{r\to \infty} e^{iqr} \langle X_j X_{j+r}^\dagger \rangle &= e^{-1 / (2\sqrt{|\kappa|})}, \qquad q = 0, \pi.
    \end{aligned}
\end{equation}
Both correlations approach nonzero constants at long distances within the harmonic approximation, explaining the tendency toward coexistence of $X$ and $Z$ orders, particularly at large $N$.

\SMsubsection{From Lifshitz theory to the dipolar Luttinger liquid}{app:Lifshitz-to-dLL}

We first derive the continuum limit of the harmonic dipolar rotor theory and then examine how the discreteness of the rotor number modifies its low-energy behavior. We introduce continuum fields $\rho(x)$ and $\Theta(x)$ for the coarse-grained rotor number $m_j$ and phase fluctuation $\varphi_j$, respectively. Defining the dual field $\Phi(x)$ through
\begin{align} \label{app:rho_Phi}
    \rho(x) = \rho_0 + \frac{\partial_x \Phi(x)}{\pi}, \quad
    [\Phi(x), \partial_y \Theta(y)] = i\pi\delta(x-y),
\end{align}
where $\rho_0$ is a uniform reference density. In the continuum limit, we have
\begin{equation}
    \begin{aligned}
        m_{j+1} - m_j &\,\longrightarrow\, \partial_x \rho(x) = \frac{1}{\pi} \partial_x^2 \Phi(x), \\
        \varphi_{j+1} - \varphi_j &\,\longrightarrow\, \partial_x \Theta(x).
    \end{aligned}
\end{equation}
Replacing the lattice sum by a spatial integral, the harmonic Hamiltonian becomes
\begin{align}
    H_{z = 2} = \int dx \left[ \frac{1}{\pi^2} (\partial_x^2 \Phi)^2 + |\kappa| (\partial_x \Theta)^2 \right],
\end{align}
which is a Lifshitz theory with dynamical exponent $z = 2$, whose low-energy dispersion is $\omega(k) = 2\sqrt{|\kappa|} k^2$.

We now incorporate the discreteness of the rotor number through a periodic potential for $\rho(x)$. Under the global $\Z_N$ symmetry generated by $U_X$, the rotor configurations transform as $m_j \to m_j+1$ (mod $N$), corresponding to $\rho \to \rho + 1$, while $U_Z$ leaves $\rho$ unchanged. Thus, the lowest-harmonic periodic potential allowed by these symmetries is
\begin{align} \label{eq:commen_pot}
    V(\rho) = -y \cos(2\pi\rho).
\end{align}
Unlike the integer-valued microscopic rotor number $m_j$, $\rho(x)$ is a continuous coarse-grained field, so this potential need not equal unity. We consider $y > 0$, which favors an integer-commensurate background $\rho_0 \in \Z$. Expanding around this background gives
\begin{align}
    V(\rho) = V(\rho_0) + \frac 12 V''(\rho_0) (\rho - \rho_0)^2 + \cdots.
\end{align}
Using Eq.~\eqref{app:rho_Phi}, this produces
$2g (\partial_x \Phi)^2$, where
\begin{align}
    g = \frac{V''(\rho_0)}{4\pi^2} > 0.
\end{align}

The resulting long-wavelength Hamiltonian is therefore
\begin{align} \label{eq:perturbed-Lifshitz}
    H_{\rm eff} = \int dx \left[ \frac{1}{\pi^2} (\partial_x^2 \Phi)^2 + |\kappa| (\partial_x \Theta)^2 + 2g (\partial_x \Phi)^2 \right].
\end{align}
Using the canonical momentum $\Pi = \partial_x \Theta / \pi$, which satisfies $[\Phi(x), \Pi(y)] = i\delta(x - y)$, the Hamiltonian for each Fourier mode becomes
\begin{align}
    H_k = \pi^2 |\kappa| \Pi_k \Pi_{-k} + \left( \frac{k^4}{\pi^2} + 2gk^2 \right) \Phi_k \Phi_{-k}.
\end{align}
The corresponding dispersion relation is
\begin{align}
    \omega^2 (k) = 4|\kappa| k^4 + 8\pi^2 |\kappa| g k^2.
\end{align}
For $g > 0$, the $k^2$ term dominates at long wavelengths, giving a linear dispersion
\begin{align}
    \omega(k) \simeq v|k|, \qquad v = \pi \sqrt{8|\kappa| g}.
\end{align}
Thus, the $g$ term is a relevant perturbation to the $z = 2$ Lifshitz theory, driving a crossover to a $z = 1$ Luttinger liquid in which the $(\partial_x^2 \Phi)^2$ term becomes irrelevant. Dropping the $(\partial_x^2 \Phi)^2$ term and matching Eq.~\eqref{eq:perturbed-Lifshitz} to
the standard Luttinger-liquid Hamiltonian form, we obtain
\begin{align}
    H_{\rm LL} = \frac{v}{2\pi} \int dx \left[ K (\partial_x \Theta)^2 + \frac 1K (\partial_x \Phi)^2 \right],
\end{align}
with
\begin{align}
    v = \pi \sqrt{8|\kappa| g}, \qquad K = \sqrt{\frac{|\kappa|}{2g}}.
\end{align}

\SMsubsection{Pad\'e estimate of the single-ordering boundary}{app:pade_SO_bd}

We estimate the large-$N$ boundary of the $Z_0$ single-order phase from the integer-rotor Hamiltonian $H(\kappa)$ in Eq.~\eqref{eq:dqr}. Writing $T_j = e^{i(\phi_{j+1} - \phi_j)}$, we have $H(\kappa) = H_0 + \kappa V$, with
\begin{align}
    H_0 = \sum_j (m_{j+1} - m_j)^2, \qquad V= -\sum_j (T_j + T_j^\dagger).
\end{align}

\SMsubsubsection{Charge gap and uniform $X$ correlations}{app:charge_gap_X_corr}

In the integer-rotor representation, $X_j$ increases $m_j$ by one, changing the total rotor number $M = \sum_j m_j$ by one. Since $H(\kappa)$ conserves $M$, we can compare the lowest energies $E_M (\kappa)$ in different sectors. In particular, define
\begin{align}
    \Delta_1 (\kappa) = E_1(\kappa) - E_0(\kappa),
\end{align}
which equals $2$ at $\kappa = 0$. We show below that uniform $X$ long-range order requires $\Delta_1$ to vanish in the thermodynamic limit. This motivates estimating the instability of the $Z_0$ single-order phase from the closing of $\Delta_1$.

Let $| \Omega_0 \rangle$ be a ground state in the $M = 0$ sector, and denote its expectation values by $\langle \cdot \rangle_0$. The operator $X_{\rm tot} = \sum_{j=1}^L X_j$ probes uniform $X$ correlations and satisfies $[M, X_{\rm tot}] = X_{\rm tot}$. Thus, $X_{\rm tot} | \Omega_0 \rangle$ and $X_{\rm tot}^\dagger | \Omega_0 \rangle$ belong to the $M = 1$ and $M = -1$ sectors, respectively. The variational principle then gives
\begin{equation}
    \begin{aligned}
        \langle X_{\rm tot}^\dagger (H - E_0) X_{\rm tot} \rangle_0 &\geq \Delta_1 \langle X_{\rm tot}^\dagger X_{\rm tot} \rangle_0, \\
        \langle X_{\rm tot} (H - E_0) X_{\rm tot}^\dagger \rangle_0 &\geq \Delta_{-1} \langle X_{\rm tot} X_{\rm tot}^\dagger \rangle_0.
    \end{aligned}
\end{equation}
Charge inversion $m_j \to -m_j$ implies $\Delta_1 = \Delta_{-1}$. Moreover, since $[X_{\rm tot}, X_{\rm tot}^\dagger] = 0$, adding the two inequalities gives~\cite{tasaki21offdiagonal}
\begin{align} \label{eq:double_comm_ineq}
    2\Delta_1 \langle X_{\rm tot}^\dagger X_{\rm tot} \rangle_0 \leq \langle [X_{\rm tot}^\dagger, [H, X_{\rm tot}]] \rangle_0.
\end{align}

The hopping term commutes with $X_{\rm tot}$. Using $n_j = m_{j+1} - m_j$, $[n_j, X_j] = -X_j$ and $[n_j, X_{j+1}] = X_{j+1}$, we obtain
\begin{align}
    [X_{\rm tot}^\dagger, [n_j^2, X_{\rm tot}]] = 4 - 2(X_j^\dagger X_{j+1} + X_j X_{j+1}^\dagger).
\end{align}
Since $X_j^\dagger X_{j+1}$ is unitary, each $[X_{\rm tot}^\dagger, [n_j^2, X_{\rm tot}]]$ is nonnegative and bounded above by $8\mathbbm{1}$. Summing over the $L$ bonds then gives
\begin{align}
    0 \preceq [X_{\rm tot}^\dagger, [H, X_{\rm tot}]] \preceq (8L) \mathbbm{1}.
\end{align}
Combining this with Eq.~\eqref{eq:double_comm_ineq}, we arrive at the charge-gap bound
\begin{align}
    \Delta_1 (\kappa) \leq \frac{4L}{\langle X_{\rm tot}^\dagger X_{\rm tot} \rangle_0}.
    \label{charge_gap_ODLRO}
\end{align}

Since $\langle X_{\rm tot}^\dagger X_{\rm tot} \rangle_0 = \sum_{i,j} \langle X_i^\dagger X_j \rangle_0$, uniform $X$ long-range order implies $\langle X_{\rm tot}^\dagger X_{\rm tot} \rangle_0 \propto L^2$ and hence $\Delta_1(\kappa) \to 0$ as $L \to \infty$. The converse does not necessarily hold: a vanishing charge gap is also compatible with algebraically decaying $X$ correlations. We therefore use the closing of $\Delta_1$ to estimate the instability of the $Z_0$ single-order phase, rather than the onset of $X$ long-range order.

\SMsubsubsection{Perturbative estimate of $\kappa_c$}{app:pert_estm_kappac}

We estimate the closing of $\Delta_1$ using the integer-rotor Hamiltonian $H(\kappa)$ obtained near the $Z_0$-ordered point $\theta = \pi/2$, which corresponds to $\kappa = 0$.

For $\kappa > 0$, the off-diagonal matrix elements of $H(\kappa)$ in the $m$ basis are nonpositive, since $V$ transfers rotor number between neighboring sites with amplitude $-\kappa$. Consider a normalized state in a fixed-$M$ sector, $| \psi \rangle = \sum_{\mathbf{m}} c_{\mathbf{m}} | \mathbf{m} \rangle$. Since $\mathrm{Re} [c_{\mathbf{m}}^* c_{\mathbf{m}'}] \leq |c_{\mathbf{m}}| |c_{\mathbf{m}'}|$, a normalized state $| \widetilde{\psi} \rangle = \sum_{\mathbf{m}} |c_{\mathbf{m}}| | \mathbf{m} \rangle$ cannot have energy greater than that of $| \psi \rangle$. A ground state $| \psi_M \rangle$ in each fixed-$M$ sector can therefore be chosen with nonnegative coefficients, $\psi_{\mathbf{m}} \geq 0$.

Let $\mathsf{T}$ denote the translation by one site and define the projector onto zero-momentum sector by $P_0 = L^{-1} \sum_{j=0}^{L-1} \mathsf{T}^j$. Since all coefficients of $| \psi_M \rangle$ are nonnegative,
\begin{align}
    \langle \psi_M | P_0 | \psi_M \rangle &= \frac 1L \sum_{j=0}^{L-1} \sum_{\mathbf{m}} \psi_{\mathbf{m}} \psi_{\mathsf{T}^{-j} \mathbf{m}} \geq \frac 1L \sum_{\mathbf{m}} \psi_{\mathbf{m}}^2 = \frac 1L > 0.
\end{align}
Thus, the ground state has a nonzero zero-momentum component. Since $[H, P_0] = 0$, we can conclude that the lowest energy in each fixed-$M$ sector is attained at zero momentum.

At $\kappa = 0$, the $M = 0$ ground state is the uniform configuration $m_j = 0$, with energy zero. In the $M = 1$ sector, the lowest-energy configurations have a single $m_j = 1$, whose unique zero-momentum superposition has energy two.  We therefore apply nondegenerate Rayleigh--Schr\"odinger perturbation theory separately in the zero-momentum $M = 0$ and $M = 1$ sectors. Expanding the energy and ground state in each $M$-sector as
\begin{align}
    E_M (\kappa) = \sum_{n=0}^\infty \kappa^n E_M^{(n)}, \qquad | \psi_M (\kappa) \rangle = \sum_{n=0}^\infty \kappa^n | \psi_M^{(n)} \rangle,
\end{align}
with $\langle \psi_M^{(0)} | \psi_M^{(n)} \rangle = 0$ for $n\geq 1$, the coefficients are obtained recursively as
\begin{equation}
    \begin{aligned}
        E_M^{(n)} &= \langle \psi_M^{(0)} | V | \psi_M^{(n-1)} \rangle, \\
        | \psi_M^{(n)} \rangle &= \frac{Q_M}{E_M^{(0)} - H_0} \left[ V | \psi_M^{(n-1)} \rangle - \sum_{i=1}^n E_M^{(i)} | \psi_M^{(n-i)} \rangle \right],
    \end{aligned}
\end{equation}
where $Q_M = 1 - | \psi_M^{(0)} \rangle \langle \psi_M^{(0)} |$, and the inverse is understood on the range of $Q_M$ within the zero-momentum sector.

Subtracting the resulting energy series gives the charge gap $\Delta_1(\kappa) = E_1 (\kappa) - E_0 (\kappa)$. The expansion coefficients up to eighth order have converged for $L\geq 10$, yielding
\begin{equation} \label{Delta_series}
    \begin{aligned}
        \Delta_1(\kappa) &= 2 - 2\kappa + \frac 16 \kappa^2 + \frac{1}{24} \kappa^3 \\
        & -\frac{33127}{665280} \kappa^4 + \frac{470812891}{4516300800} \kappa^5 \\
        & -0.135343937856 \kappa^6 + 0.199759095251 \kappa^7 \\
        & -0.283672169995 \kappa^8 + O(\kappa^9).
    \end{aligned}
\end{equation}
This truncated series is reliable near $\kappa = 0$. To estimate the closing of $\Delta_1 (\kappa)$, we construct Pad\'e approximants $[p/q] = P_p (\kappa) / Q_q (\kappa)$ with $p + q = 8$, where $P_p$ and $Q_q$ are polynomials of degree at most $p$ and $q$, respectively, chosen to reproduce the series through eighth order. We then find the smallest positive real zero of $P_p(\kappa)$. Table~\ref{tab:pade_zeros} lists the zeros and poles obtained from Eq.~\eqref{Delta_series}. No listed approximant has a positive pole before its first positive zero. The near-diagonal approximants $[3/5]$, $[4/4]$, and $[5/3]$ give $1.149374$, $1.148610$, and $1.149677$, consistent with the estimate
\begin{align}
    \kappa_c \simeq 1.1491(7).
\end{align}

\begin{table}[t]
    \centering
    \begin{tabular}{ccc}
        \hline\hline
        $[p/q]$ & $\kappa_{\rm zero}$ & $\kappa_{\rm pole}$\\
        \hline
        $[1/7]$ & 0.961831 & 1.062282\\
        $[2/6]$ & 1.143439 & --\\
        $[3/5]$ & 1.149374 & 2.575621\\
        $[4/4]$ & 1.148610 & --\\
        $[5/3]$ & 1.149677 & 2.377296\\
        $[6/2]$ & 1.149678 & 2.377085\\
        $[7/1]$ & 1.146407 & --\\
        \hline\hline
    \end{tabular}
    \caption{Smallest positive real zeros and poles of the eighth-order Pad\'e approximants to Eq.~\eqref{Delta_series}. A dash denotes no positive real pole.}
    \label{tab:pade_zeros}
\end{table}

Using $\kappa = \cot\theta / (2\pi/N)^2$, the resulting estimate of the large-$N$ boundary of the SO $Z_0$ phase is
\begin{align} \label{theta_cZ}
    \theta_{c,Z} = \frac{\pi}{2} - \arctan \left( \frac{4\pi^2 \kappa_c}{N^2} \right) \simeq \frac{\pi}{2} - \frac{4\pi^2 \kappa_c}{N^2}.
\end{align}By $X\leftrightarrow Z$ duality, the corresponding boundary of the SO $X_0$ phase is
\begin{align} \label{theta_cX}
    \theta_{c,X} = \arctan \left( \frac{4\pi^2 \kappa_c}{N^2} \right) \simeq \frac{4\pi^2 \kappa_c}{N^2}.
\end{align}
We define $N_*$ as the value at which the two single-order instability boundaries meet at the self-dual point:
\begin{align}
    \theta_{c,X} (N_*) = \theta_{c,Z} (N_*) = \frac{\pi}{4}, \qquad N_* = 2\pi \sqrt{\kappa_c} \simeq 6.735,
\end{align}
which estimates the onset of the coexistence phase. For $N > N_*$, the two boundaries satisfy $\theta_{c,X} < \pi/4 < \theta_{c,Z}$, leaving an intervening region between the two SO phases in which the C phase can emerge. This is a large-$N$ estimate, yet the numerical observation that C order appears for $N \geq 6$ is qualitatively consistent with this prediction.

At finite $N$, however, the intervening region need not be entirely occupied by the C phase. Indeed, dLL$_{\rm I}$ phases are numerically observed for $N\geq 6$, consistent with the dipolar rotor theory when the discreteness of the rotor number is taken into account (see Sec.~\ref{app:Lifshitz-to-dLL}). Notably, the dLL$_{\rm I}$ regions become narrower with increasing $N$. The Pad\'e analysis captures the instability of the SO phase but does not by itself determine the width or the two boundaries of the intervening dLL. Therefore, the observed narrowing of the dLL$_{\rm I}$ phases suggests that the estimates in Eqs.~\eqref{theta_cZ} and \eqref{theta_cX} increasingly approach the C/SO boundary as $N\to \infty$.

\SMsection{\texorpdfstring{Dipolar Luttinger liquid for odd $N$ near $\theta = \pi$}{Dipolar Luttinger liquid for odd N near theta=pi}}{app:even-odd}

For odd $N$, no integer power of $\omega$ equals $-1$, so the staggered clock rotation used for even $N$ is not available. Near $\theta = \pi$, the key distinction is that each bond has two degenerate minima, giving an exponentially degenerate ground-state (EDGS) manifold at $\theta = \pi$. Here we derive the effective Hamiltonian acting within this EDGS manifold and show how its mobile sector reduces to an XXZ chain, yielding a dipolar Luttinger liquid (dLL) phase near $\theta = \pi$. We then discuss how corrections beyond this effective theory can terminate the dLL phase through a BKT transition.

\SMsubsection{Projection onto the EDGS manifold}{app:oddN-boundary-layer}

Using $\delta = \theta - \pi$ and a bond variable $l_j = s_{j+1} - s_j \pmod N$, the exact QTC Hamiltonian can be written as $H_{\rm QTC} = H_0 + V$, with
\begin{align}
    H_0 = 2\cos\delta \sum_j \cos(\epsilon l_j), \quad V = \sin\delta \sum_j (T_j + T_j^\dagger),
\end{align}
where $\epsilon = 2\pi / N$ and $T_j = Z_j Z_{j+1}^\dagger$. At $\delta = 0$, each bond has two degenerate minima $l_j = \pm N'$ with $N' = (N-1)/2$, each of bond energy $-2\cos(\pi/N)$. The next lowest bond energy is $-2\cos(3\pi/N)$, so the local energy cost of
leaving the minimum is
\begin{align}
    \Delta_{\theta = \pi} = 2\left( \cos\frac{\pi}{N} - \cos\frac{3\pi}{N} \right) = 4\sin\frac{\pi}{N} \sin\frac{2\pi}{N}.
\end{align}
For an open chain, this is the excitation gap above the EDGS manifold. On a periodic ring, the bond variables additionally satisfy $\sum_j l_j = 0 \pmod N$. This constraint can always be satisfied within the EDGS manifold for $L\geq N$. Since we are interested in the thermodynamic limit, we assume $L\geq N$ throughout. The local scale $\Delta_{\theta = \pi}$ sets the energy cost of the virtual excitations entering the effective Hamiltonian below.

Let $P$ denote the projector onto the EDGS manifold and $Q = 1 - P$. The unperturbed ground-state energy on a ring is $E_0 = -2L\cos\delta \cos(\pi/N)$. Acting with $T_j$ changes three consecutive bond variables according to
\begin{align}
    (\delta l_{j-1}, \delta l_j, \delta l_{j+1}) = (1, -2, 1),
\end{align}
with the opposite change for $T_j^\dagger$. For odd $N \geq 5$, the central bond variable $\delta l_j$ leaves the set $\{+N',-N'\}$ and hence $PVP = 0$. This is the key distinction from $N = 3$, where the projected first-order dynamics is nonzero and produces the two-component $c = 2$ liquid of Ref.~\cite{lee12quantum}. For odd $N \geq 5$, the leading effective Hamiltonian arises at second order:
\begin{align}
H_{\rm eff}^{(2)} = -PVQ [Q (H_0-E_0) Q]^{-1} QVP.
\end{align}

\SMsubsection{Second-order effective Hamiltonian}{app:second-order-eff-hamiltonian}

Within the EDGS manifold, each bond is restricted to the two states $l_j = \pm N'$. We represent this by a pseudospin:
\begin{equation}
    \begin{aligned}
    | 0 \rangle_j &\equiv | {+}N' \rangle_j, \qquad \sigma_j^z | 0 \rangle_j = | 0 \rangle_j, \\
    | 1 \rangle_j &\equiv | {-}N' \rangle_j, \qquad \sigma_j^z | 1 \rangle_j = -| 1 \rangle_j.
    \end{aligned}
\end{equation}

Since $PVP=0$, the leading dynamics within the EDGS manifold arises at second order. The only two-step process connecting two different states in the EDGS manifold is
\begin{align}
    (N', -N', -N', N') \,\,\leftrightarrow\,\, (-N', N', N', -N'),
\end{align}
or equivalently, $0110 \leftrightarrow 1001$. In the pseudospin language, this process is generated by 
\begin{align}
    \sigma_{j-1}^{+} \sigma_j^{-} \sigma_{j+1}^{-} \sigma_{j+2}^{+} + \mathrm{h.c.},
\end{align}
with $\sigma_j^\pm = (\sigma_j^x \pm i \sigma_j^y) / 2$. This transition can occur in two orders, $T_j T_{j+1}$ and $T_{j+1} T_j$, and each intermediate state has excitation energy $2\cos\delta\, \Delta_{\theta=\pi}$. Summing the two paths gives the effective hopping amplitude $-(\sin^2 \delta / \cos\delta) t_N$, with
\begin{align}
    t_N \equiv \frac{1}{\Delta_{\theta=\pi}} = \frac{1}{2(c_1 - c_3)},
\end{align}
where $c_r = \cos(r\pi/N)$. 

Second-order processes can also return to the same state within the EDGS manifold and produce diagonal energy shifts. Up to reflection and simultaneous reversal of all signs, there are three inequivalent local bond-variable patterns:
\begin{align}
    (+++), \qquad (++-), \qquad (+-+),
\end{align}
with the corresponding amplitudes
\begin{align}
    F_1 &= \frac{1}{2(c_1 - c_5)} + \frac{1}{6(c_1 - c_3)}, \nonumber \\
    F_2 &= \frac{1}{2(2c_1 - c_3 - c_5)} + \frac{1}{4(c_1 - c_3)}, \label{F1F2F3} \\
    F_3 &= \frac{1}{2(c_1 - c_3)} + \frac{1}{2(3c_1 - 2c_3 - c_5)}. \nonumber
\end{align}
Each term in Eq.~\eqref{F1F2F3} is the inverse excitation energy of the intermediate state reached by one of the two possible first virtual hops, $T_j$ or $T_j^\dagger$. Writing $\tau_j = \pm 1$ for the eigenvalue of $\sigma_j^z$, the corresponding three-bond energy shift can be written as
\begin{align}
    f(\tau_{j-1}, \tau_j, \tau_{j+1}) = a_0 + a_1 (\tau_{j-1} \tau_j + \tau_j \tau_{j+1}) + a_2 \tau_{j-1} \tau_{j+1},
\end{align}
with
\begin{align}
    a_0 = \frac{F_1 + F_3 + 2F_2}{4}, \quad a_1 = \frac{F_1 - F_3}{4}, \quad a_2 = \frac{F_1 + F_3 - 2F_2}{4}.
\end{align}

Combining the hopping and diagonal shift terms, the second-order effective Hamiltonian near $\theta = \pi$ becomes
\begin{equation}
    \begin{aligned}
    H_{\rm eff}^{(2)} &= E_0^{(2)} - t_N \frac{\sin^2 \delta}{\cos\delta} \sum_j \Big[ J_{1,N} \sigma_j^z \sigma_{j+1}^z + J_{2,N} \sigma_j^z \sigma_{j+2}^z \\
    &\hspace{80pt} + \sigma_{j-1}^{+} \sigma_j^{-} \sigma_{j+1}^{-} \sigma_{j+2}^{+} + \mathrm{h.c.} \Big],
    \end{aligned} \label{Heff2}
\end{equation}
where 
\begin{align}
E_0^{(2)} = -La_0 \frac{\sin^2 \delta}{\cos\delta}
\end{align}
and
\begin{equation}
    \begin{aligned}
    J_{1,N} &= \frac{2a_1}{t_N} = -\frac{(\alpha_N - 1)(\alpha_N + 3)}{3\alpha_N (\alpha_N + 2)}, \\
    J_{2,N} &= \frac{a_2}{t_N} = \frac{(\alpha_N + 3)(\alpha_N^2 + 2)}
    {12\alpha_N (\alpha_N + 1)(\alpha_N + 2)},
    \end{aligned}
\end{equation}
with $\alpha_N = 1 + 2 \cos (2\pi/N)$.

The pair-hopping process $0110 \leftrightarrow 1001$ preserves both the total charge $Q = \sum_j \sigma_j^z$ and the bulk dipole moment $D = \sum_j j \sigma_j^z$. Thus, the dynamics under Eq.~\eqref{Heff2} cannot connect configurations with different $(Q, D)$ values. Moreover, because the only allowed off-diagonal move is the local resonance $0110 \leftrightarrow 1001$, even a fixed $(Q, D)$ sector can further decompose into multiple disconnected Krylov sectors. In particular, configurations containing no $0110$ or $1001$ pattern admit no allowed pair hop and therefore form frozen sectors.

\SMsubsection{Mobile sector and the XXZ Luttinger liquid}{app:XXZ-LL} 

We next focus on a mobile invariant sector of the second-order effective Hamiltonian $H_{\rm eff}^{(2)}$ and show that its dynamics reduces to an XXZ chain with a Luttinger liquid phase.

Recall that $| 0 \rangle_j \equiv | {+}N' \rangle_j$ and $| 1 \rangle_j \equiv | {-}N' \rangle_j$. For even $L$, choose one of the two block origins $a = 0, 1$ and group the bond variables as $(l_{2j+a}, l_{2j+a+1})$, restricting each block to the two pseudospin configurations $01$ and $10$. This subspace is invariant under the pair-hopping process because $01|10 \leftrightarrow 10|01$. Since $s_{j+2} - s_j = l_{j+1} + l_j \pmod N$, the block constraint $l_{2j+a} + l_{2j+a+1} = 0$ is equivalently
\begin{align}
    s_{2j+a+2} = s_{2j+a} \pmod N,
\end{align}
for a fixed choice of $a = 0$ or $1$. Namely, one of the two sublattices of the original variables $s_j$ is fixed throughout the mobile sector.  
 
We henceforth choose $a = 0$ without loss of generality and identify
\begin{align}
    | \uparrow \rangle_{j'} = | 01 \rangle_{2j,2j+1}, \qquad | \downarrow \rangle_{j'} = | 10 \rangle_{2j,2j+1}.
\end{align}
The resonance $01|10 \leftrightarrow 10|01$ then becomes the nearest-neighbor exchange of the block spins. Denoting the corresponding spin-$1/2$ operators by ${\bf S}_{j'}$, we have within each block
\begin{align}
    \sigma^z_{2j} = 2S^z_{j'}, \qquad \sigma^z_{2j+1} = -2S^z_{j'}.
\end{align}
Substituting these identifications into Eq.~\eqref{Heff2}, the $\sigma_j^z \sigma_{j+1}^z$ and $\sigma_j^z \sigma_{j+2}^z$ interactions reduce to nearest-neighbor terms $S_{j'}^z S_{j'+1}^z$ between the block spins, while the pair-hopping term becomes $S_{j'}^- S_{j'+1}^+ + S_{j'}^+ S_{j'+1}^- = 2(S_{j'}^x S_{j'+1}^x + S_{j'}^y S_{j'+1}^y)$. After the staggered rotation $S_{j'}^\pm \rightarrow (-1)^{j'} S_{j'}^\pm$, Eq.~\eqref{Heff2} reduces, up to a constant and a possible boundary twist, to 
\begin{align}
    H_{\rm XXZ}^{(N \geq 5)} &= 2t_N \frac{\sin^2 \delta}{\cos\delta} \sum_{j'} \left( S_{j'}^x S_{j'+1}^x + S_{j'}^y S_{j'+1}^y + \Delta_N S_{j'}^z S_{j'+1}^z \right), \nn
    \Delta_N &= 2J_{1,N} - 4J_{2,N} = -\frac{\alpha_N (\alpha_N + 3)}{(\alpha_N + 1)(\alpha_N + 2)}.
    \label{app:oddN-XXZ}
\end{align}
For a periodic ring, the staggered rotation leaves a twist in the closing bond when the block chain has odd length. We therefore take $L = 0\pmod 4$ when discussing finite periodic systems; this boundary twist does not affect the bulk thermodynamic description.

For odd $N\geq 5$, the anisotropy parameter $\Delta_N$ satisfies $-1 < \Delta_N < 1$. Thus, the mobile-sector XXZ Hamiltonian Eq.~\eqref{app:oddN-XXZ} lies in the critical regime, with a zero magnetization ground state described by a $c=1$ Luttinger liquid. The corresponding Luttinger parameter and velocity are
\begin{align}
    K_N = \frac{\pi}{2(\pi - \arccos \Delta_N)}, \qquad v_N \propto \frac{\sin^2 \delta}{\cos\delta},
    \label{app:oddN-LL}
\end{align}
respectively. For example, $\Delta_5 \simeq -0.7889$ and $K_5 \simeq 2.373$, while $\Delta_N \to -9/10$ as $N\to \infty$.

The Luttinger mode has a simple interpretation in the original $N$-clock variables. The XXZ Hamiltonian Eq.~\eqref{app:oddN-XXZ} conserves the block-spin magnetization $\sum_{j'} S_{j'}^z$. Using $\sigma^z_{2j} = 2S^z_{j'}$ and $\sigma^z_{2j+1} = -2S^z_{j'}$, the dipole moment within the EDGS manifold can be written as
\begin{align}
    \sum_j j\sigma_j^z &= \sum_{j'} \left[ 2j' \sigma^z_{2j'} + (2j' + 1) \sigma^z_{2j'+1} \right] = -2\sum_{j'} S_{j'}^z.
\end{align}
Thus, the conserved magnetization of the effective XXZ chain is directly inherited from the dipole moment of the bond variables $l_j$, showing that the propagating Luttinger mode is dipolar and charge-neutral. Such a phase is referred to as a dipolar Luttinger liquid (dLL).

The same block structure also explains the nearly staggered $X$ background near $\theta = \pi$. Consider the mobile sector with the even sublattice fixed to $s_{2j} = c$. Since each block is either $01$ or $10$, the odd sublattice takes $s_{2j+1} = c\pm N' \pmod N$, and therefore $X_{2j} = e^{-i\epsilon c}$ and $X_{2j+1} = -e^{-i\epsilon c} e^{\pm i\pi/N}$. For the zero-magnetization XXZ ground state, $\langle S_{j'}^z \rangle = 0$, so the two block orientations $01$ and $10$ have equal local expectation values. Averaging the two values gives $\langle X_{2j+1} \rangle = -e^{-i\epsilon c} \cos(\pi/N)$. The corresponding uniform and staggered order parameters are therefore
\begin{align}
    m_{X,0} = \frac{1 - \cos(\pi/N)}{2}, \qquad m_{X,\pi} = \frac{1 + \cos(\pi/N)}{2}.
    \label{XXZ_mX}
\end{align}
Therefore, the dipolar Luttinger mode can coexist with strong staggered $X$ order and weaker uniform $X$ order. For $N = 5$, Eq.~\eqref{XXZ_mX} predicts $m_{X,0} \simeq 0.0955$ and $m_{X,\pi} \simeq 0.905$, which are in good agreement with Fig.~\ref{fig:dLL_near_pi}(a) of the main text.

The EDGS projection is controlled when the perturbative scale remains small compared with the gap out of the EDGS manifold:
\begin{align}
    |\tan\delta| \ll \Delta_{\theta=\pi} = 4\sin\frac{\pi}{N} \sin\frac{2\pi}{N} \sim N^{-2}.
\end{align}
This condition sets only the validity scale of the perturbation theory, and the XXZ mapping Eq.~\eqref{app:oddN-XXZ} establishes a $c = 1$ mobile invariant sector of the leading effective Hamiltonian, rather than proving that it is the global ground state among all fragmented sectors. Nevertheless, since the dLL originates from this sector, the loss of parametric separation when $|\tan\delta| \simeq \Delta_{\theta=\pi}$ suggests that the width of the dLL phase scales as $|\delta| = O(N^{-2})$ for large odd $N$.

\SMsubsection{\texorpdfstring{Fourth-order effective Hamiltonian for $N = 5$}{Fourth-order effective Hamiltonian for N = 5}}{app:N5-fourth-order}

The second-order effective Hamiltonian fixes the XXZ anisotropy and therefore Luttinger parameter $K_N$ independent of $\delta$. The leading dependence of $K$ on $\delta = \theta - \pi$ arises from higher-order processes. For $N = 5$, writing $t = \tan\delta$, the fourth-order effective Hamiltonian can be written, up to a
constant energy shift, as
\begin{align}
    H_{\rm eff}^{(4)} = \frac{\sin^2 \delta}{\sqrt{5} \cos\delta} \left[F_2 + t F_3 + t^2 \left( F_4^{({\rm c})} + F_4^{({\rm nc})} \right) \right].
    \label{app:oddN-full4}
\end{align}
Here, the second-order contribution is
\begin{equation}
    \begin{aligned}
    F_2 &= -\sum_j \Big[ J_1 \sigma_j^z \sigma_{j+1}^z + J_2 \sigma_j^z \sigma_{j+2}^z \\
    &\hspace{35pt} + \sigma_j^- \sigma_{j+1}^+ \sigma_{j+2}^+ \sigma_{j+3}^- + \mathrm{h.c.} \Big],
    \end{aligned}
\end{equation}
while the third-order contribution is
\begin{align}
    F_3 = c_3 \sum_j \left( \sigma_j^- \sigma_{j+1}^+ \sigma_{j+3}^+ \sigma_{j+4}^- + \mathrm{h.c.} \right).
\end{align}
The numerical values of all coefficients in $H_{\rm eff}^{(4)}$ are summarized in Table~\ref{tab:N5eff}. At fourth order, the effective Hamiltonian separates into terms that conserve both the emergent charge $Q = \sum_j \sigma_j^z$ and the dipole moment $D = \sum_j j \sigma_j^z$, and terms that preserve the microscopic $\mathbb Z_5$ symmetry but violate at least one of these emergent conservation laws. The charge- and dipole-conserving part is
\begin{align}
    F_4^{({\rm c})} &= \sum_j \Big[ a_1 \sigma_j^z \sigma_{j+1}^z + a_2 \sigma_j^z \sigma_{j+2}^z + a_3 \sigma_j^z \sigma_{j+3}^z + a_4 \sigma_j^z \sigma_{j+4}^z \Big]
    \nonumber \\
    &\hspace{10pt} + \sum_j \Big[ b_1 \sigma_j^z \sigma_{j+1}^z \sigma_{j+2}^z \sigma_{j+3}^z + b_2 \sigma_j^z \sigma_{j+1}^z \sigma_{j+3}^z \sigma_{j+4}^z \nonumber \\
    &\hspace{35pt} + b_3 \left( \sigma_j^z \sigma_{j+2}^z \sigma_{j+3}^z \sigma_{j+4}^z + \sigma_j^z \sigma_{j+1}^z \sigma_{j+2}^z \sigma_{j+4}^z \right) \Big] \nonumber \\
    &\hspace{10pt} + \sum_j \Big[ h_1 \sigma_j^- \sigma_{j+1}^+ \sigma_{j+2}^+ \sigma_{j+3}^- + h_2 \sigma_j^- \sigma_{j+1}^+ \sigma_{j+4}^+ \sigma_{j+5}^- \nonumber \\
    &\hspace{35pt} + h_3 \left( \sigma_j^- \sigma_{j+1}^+ \sigma_{j+2}^+ \sigma_{j+3}^- \sigma_{j+4}^z \sigma_{j+5}^z \right. \\
    &\hspace{90pt} \left. + \sigma_j^z \sigma_{j+1}^z \sigma_{j+2}^- \sigma_{j+3}^+ \sigma_{j+4}^+ \sigma_{j+5}^- \right) \nonumber \\
    &\hspace{35pt} + h_4 \sigma_j^- \sigma_{j+1}^+ \sigma_{j+2}^z \sigma_{j+3}^z \sigma_{j+4}^+ \sigma_{j+5}^- \nonumber \\
    &\hspace{35pt} + h_5 \sigma_j^- \sigma_{j+1}^+ \sigma_{j+2}^+ \sigma_{j+4}^- \sigma_{j+5}^- \sigma_{j+6}^+ + \mathrm{h.c.} \Big]. \nonumber
\end{align}
The remaining non-conserving fourth-order terms are
\begin{align}
    F_4^{({\rm nc})} &= \sum_j \Big[ d \big( \sigma_j^- \sigma_{j+2}^- \sigma_{j+3}^+ \sigma_{j+4}^+ + \sigma_j^- \sigma_{j+1}^- \sigma_{j+2}^+ \sigma_{j+4}^+ \big) \nonumber \\
    &\hspace{35pt} + e \sigma_j^+ \sigma_{j+1}^+ \sigma_{j+2}^+ \sigma_{j+3}^+ \sigma_{j+4}^+ + \mathrm{h.c.} \Big].
\end{align}

\begin{table}[t]
    \centering
    \begin{tabular}{c r @{\hspace{0.4em}} | @{\hspace{0.4em}} c r @{\hspace{0.4em}} | @{\hspace{0.4em}} c r}
        \hline\hline
        $J_1$ & $-1.6251\times 10^{-1}$ & $a_1$ & $-1.2161\times 10^{-1}$ & $h_1$ & $ 2.2966\times 10^{-1}$\\
        $J_2$ & $1.1596\times 10^{-1}$ & $a_2$ & $3.3705\times 10^{-2}$ & $h_2$ & $-5.4283\times 10^{-1}$ \\
        $c_3$ & $8.6896\times 10^{-1}$ & $a_3$ & $-2.8255\times 10^{-2}$ & $h_3$ & $-8.8820\times 10^{-3}$ \\
        & & $a_4$ & $1.5802\times 10^{-2}$ & $h_4$ & $-1.8911\times 10^{-1}$ \\
        & & $b_1$ & $9.6105\times 10^{-2}$ & $h_5$ & $-2.0000\times 10^{-1}$ \\
        & & $b_2$ & $4.3314\times 10^{-5}$ & $d$ & $-1.9881\times 10^{-1}$ \\
        & & $b_3$ & $-4.6072\times 10^{-3}$ & $e$ & $-1.3897\times 10^{-1}$ \\
        \hline\hline
    \end{tabular}
    \caption{Coefficients of the fourth-order effective Hamiltonian $H_{\rm eff}^{(4)}$ near $\theta = \pi$ for $N = 5$.}
    \label{tab:N5eff}
\end{table}

To determine the leading variation of the Luttinger parameter $K_5$, we consider corrections up to order $t^2$ relative to the leading second-order Hamiltonian. Let $P_{\rm mob}$ denote the projector onto the mobile sector introduced in Sec.~\ref{app:XXZ-LL}. Since $P_{\rm mob} F_3 P_{\rm mob} = 0$, the third-order term $F_3$ contributes only through virtual excursions outside the mobile sector. Its leading effect on the mobile-sector dynamics therefore appears at relative order $t^2$. In contrast, $F_4^{({\rm c})}$ acts directly within the mobile sector and contributes at the same order. The non-conserving terms $F_4^{({\rm nc})}$ also satisfy $P_{\rm mob} F_4^{({\rm nc})} P_{\rm mob} = 0$, and hence do not directly renormalize $K_5$ at this order. Instead, they can break the emergent pseudospin $U(1)$ symmetry down to the microscopic $\mathbb Z_5$ symmetry by generating the phase-locking perturbation through virtual processes outside the mobile sector; see Sec.~\ref{app:oddN-continuum-BKT} below for the continuum description.

We therefore use the charge- and dipole-conserving effective Hamiltonian
\begin{align}
    H_{\rm eff}^{(4,{\rm c})} (\delta) = \frac{\sin^2 \delta}{\sqrt{5} \cos\delta} \left[ F_2 + t F_3 + t^2 F_4^{({\rm c})} \right]
    \label{Heff_4_conv}
\end{align}
to determine the renormalized Luttinger parameter $K_5 (\delta)$. On an open chain, this Hamiltonian conserves both $Q$ and $D$ exactly, which we exploit in the DMRG calculation.

The conserving Hamiltonians Eq.~\eqref{Heff_4_conv} at $\delta$ and $-\delta$ are related by the diagonal unitary
\begin{align} \label{Heff4c_sym}
    U = \exp \left[\frac{i\pi}{4} \sum_j j^2 \sigma_j^z \right],
\end{align}
under which $F_2$ and $F_4^{({\rm c})}$ are invariant while $F_3 \to -F_3$. Hence $U H_{\rm eff}^{(4,{\rm c})} (\delta) U^\dagger = H_{\rm eff}^{(4,{\rm c})} (-\delta)$ and $K_5 (\delta) = K_5 (-\delta)$. This is a property of the conserving effective theory and does not imply an exact reflection symmetry of the full odd-$N$ QTC once the clock-locking processes are restored.

\SMsubsection{Continuum theory and BKT transition}{app:oddN-continuum-BKT}

We now formulate the low-energy continuum theory of the mobile sector and identify how it becomes unstable upon moving away from $\theta = \pi$, eventually giving way to a gapped phase. As discussed in Sec.~\ref{app:XXZ-LL}, the mobile sector maps at leading order to a critical XXZ chain [Eq.~\eqref{app:oddN-XXZ}] with $-1 < \Delta_N < 1$. Its long-wavelength limit is therefore a $c = 1$ Luttinger liquid, described by a Gaussian fixed line parameterized by the velocity $v$ and Luttinger parameter $K$:
\begin{align}
    H_{\rm LL} = \frac{v}{2\pi} \int dx \left[ K (\partial_x \Theta)^2 + \frac 1K (\partial_x\Phi)^2 \right],
\end{align}
where the fields $\Theta$ and $\Phi$ satisfy
\begin{align}
    [\Phi(x), \partial_y \Theta(y)] = i\pi \delta(x - y), \quad \Theta \equiv \Theta + 2\pi, \quad \Phi \equiv \Phi + \pi.
\end{align}
With this normalization convention, the vertex operators have scaling dimensions
\begin{align}
    \dim \left[ e^{im\Theta} \right] = \frac{m^2}{4K}, \qquad \dim \left[ e^{i2p\Phi} \right] = p^2 K.
\end{align}

Within the mobile sector with the block origin chosen as $a = 0$, the $Z$ operator on the fluctuating odd sublattice changes the local block-spin state and projects onto a spin-lowering operator:
\begin{align}
    P_{\rm mob} Z_{2j'+1} P_{\rm mob} \propto \widetilde{S}_{j'}^-,
\end{align}
where $\widetilde{S}_{j'}^- = (-1)^{j'} S_{j'}^-$ contains a factor of $(-1)^{j'}$ reflecting the staggered rotation used in the XXZ mapping. In the continuum, we have $S_{j'}^- \sim (-1)^{j'} e^{i\Theta (x_{j'})}$ and hence $\widetilde{S}_{j'}^- \sim e^{i\Theta (x_{j'})}$. Since $U_X Z U_X^\dagger = e^{-2\pi i/N}Z$ and $P_{\rm mob} Z_{2j'+1} P_{\rm mob} \sim e^{i\Theta (x_{j'})}$, the symmetry $U_X$ acts as
\begin{align}
    \Theta \longrightarrow \Theta-\frac{2\pi}{N}.
\end{align}
A phase vertex operator $e^{im\Theta}$ is therefore invariant under the microscopic $\Z_N$ symmetry only when $m$ is an integer multiple of $N$. Consequently, the leading clock-anisotropy perturbation is $\cos(N\Theta)$.

The dual field $\Phi$ describes block-spin density fluctuations through $\delta \rho(x) \propto\partial_x \Phi(x)$. At zero block magnetization, equivalently at half-filling $\rho_0 = 1/2$, the Fermi momentum is $k_F = \pi/2$. The leading $2k_F$ density harmonic $\cos(2\Phi)$ carries momentum $\pi$ and changes sign under translation by one block, so it is forbidden as a uniform bulk perturbation. By contrast, $\cos(4\Phi)$ carries momentum $4k_F = 2\pi \equiv 0 \pmod{2\pi}$ and is the leading commensurate umklapp perturbation. 

Combining the leading clock-anisotropy and umklapp perturbations, the low-energy Hamiltonian takes the form
\begin{align}
    H = H_{\rm LL} + g_u \int dx \cos(4\Phi) + g_N \int dx \cos(N\Theta).
    \label{HLL_perturbed}
\end{align}
The two perturbations in Eq.~\eqref{HLL_perturbed} have scaling dimensions $d_u = 4K$ and $d_N = N^2 / 4K$, respectively, and the corresponding dimensionless couplings $y_u$ and $y_N$ obey the following linearized RG equations:
\begin{equation}
    \begin{aligned}
    \frac{d y_u}{d\ell} &= (2 - 4K) y_u + \cdots, \\
    \frac{d y_N}{d\ell} &= \left(2 - \frac{N^2}{4K}\right) y_N + \cdots.
    \end{aligned}
\end{equation}
Thus, both perturbations are irrelevant, and the Gaussian fixed line is
stable, when
\begin{align}
    \frac 12 < K < \frac{N^2}{8}.
    \label{stable_K_window}
\end{align}

For $N = 5$, the second-order Luttinger parameter $K_5 \simeq 2.373$ [see Eq.~\eqref{app:oddN-LL}] lies within the stable window Eq.~\eqref{stable_K_window}. The higher-order corrections derived in Sec.~\ref{app:N5-fourth-order} remove the artificial $\delta$-independence of the second-order theory and renormalize $K_5$. In particular, the fourth-order conserving effective Hamiltonian $H_{\rm eff}^{(4,{\rm c})}$ [Eq.~\eqref{Heff_4_conv}] gives $K_5 (\delta) = K_5 (-\delta)$ exactly. The numerics on the QTC indeed shows that $K_5$ is nearly symmetric about $\theta = \pi$. As $|\delta|$ increases, $K_5$ reaches the upper stability boundary $K_c = 25/8$ at
\begin{align}
    \theta_{\rm BKT}^{\pm} \simeq (1 \pm 0.0725) \pi,
\end{align}
where the clock-anisotropy perturbation $\cos(5\Theta)$ becomes marginal. Beyond this point it becomes relevant, eventually pinning $\Theta$ and opening a gap. Thus, both dLL terminations are governed by phase-locking BKT transitions. A numerical test of this identification is given in Sec.~\ref{app:O5-clock}. 

The adjacent gapped phases reached after phase locking are $X'_\pi + Z'_0$ for $\theta < \pi$ and $X'_\pi + Z'_\pi$ for $\theta > \pi$. In the thermodynamic limit, the corresponding $Z$ order parameters are expected to vanish continuously upon approaching the dLL boundaries. The alternative umklapp instability would require $K_5$ to decrease to $1/2$, for which no numerical tendency is observed.

The same low-energy mechanism is expected to apply to larger odd $N \geq 7$. The leading clock-anisotropy perturbation is $\cos(N\Theta)$, which becomes marginal at $K_c = N^2 / 8$. Thus, the dLL phase near $\theta = \pi$ is expected to terminate through the phase-locking BKT transition once the renormalized Luttinger parameter $K_N$ reaches this threshold, with an $N$-dependent phase boundary. 

A characteristic feature of the BKT transition is the exponentially large correlation length $\xi$ on the ordered side. Approaching either boundary,
\begin{align}
    \xi^\pm \sim \xi_0^\pm \exp \left[ \frac{b_\pm}{\sqrt{| \theta - \theta_{\rm BKT}^{\pm} |}} \right], \qquad \Delta_{\rm LL} \sim \frac{v}{\xi^{\pm}},
\end{align}
where $\Delta_{\rm LL}$ is the gap acquired by the neutral mode propagating in the dLL. Close to the transition, $\xi^\pm$ can be much larger than the correlation length numerically accessible at finite bond dimension $\chi$. Consequently, an MPS with finite $\chi$ can make part of the thermodynamically ordered region appear LL-like, and the corresponding $Z$ order parameters can appear to drop abruptly to zero. In the $\chi\to \infty$ limit, this crossover is expected to move toward the thermodynamic BKT boundary, where the $Z$ order vanishes continuously.

Finally, the near equality of $\theta_{\rm BKT}^{-}$ and $\theta_{\rm BKT}^{+}$ does not imply an exact reflection symmetry of the original QTC about $\theta = \pi$. The exact $\delta \leftrightarrow -\delta$ equivalence applies to the conserving effective Hamiltonian $H_{\rm eff}^{(4,{\rm c})}$, whereas the full microscopic model need not obey this relation. Thus, the long-distance Luttinger parameters and the transition locations can remain nearly symmetric, even though the microscopic signatures of phase-locking need not be symmetric on the two sides.

\SMsubsection{Microscopic probe of clock anisotropy}{app:O5-clock}

We now test the clock-anisotropy mechanism discussed in Sec.~\ref{app:oddN-continuum-BKT} directly at the original QTC. For $N = 5$, consider the operator supported on five consecutive sites of the odd sublattice
\begin{align}
    O_5(j') = \prod_{k=0}^4 Z_{2j'+2k+1},
\end{align}
which is invariant under the microscopic $\mathbb Z_5$ symmetry. Since $P_{\rm mob} O_5(j') P_{\rm mob} \propto \prod_{k=0}^4 \widetilde{S}_{j'+k}^-$ and $\widetilde{S}_{j'}^- \sim e^{i\Theta (x_{j'})}$ in the continuum, the long-wavelength operator content of $O_5$ is given by
\begin{align}
    O_5(j') = C e^{5i\Theta (\bar{x})} + \cdots,
    \label{app:O5-OPE}
\end{align}
where $C$ is a nonuniversal constant, $\bar{x}$ denotes the center of the five-site cluster, and the ellipsis denotes subleading operators allowed by the same microscopic symmetries. The leading $e^{5i\Theta}$ component of $O_5$ therefore has scaling dimension
\begin{align}
    x_5 = \frac{25}{4K},
\end{align}
and contributes to the connected correlator as
\begin{align}
    \langle O_5 (0) O_5^\dagger (r) \rangle_c \propto r^{-2x_5}.
    \label{app:O5-dimension}
\end{align}
The marginality condition $x_5 = 2$, or equivalently, $K = 25/8$ is precisely the BKT criterion derived from the effective theory discussed in Sec.~\ref{app:oddN-continuum-BKT}. Importantly, $x_5$ can be extracted directly from the $O_5$ correlator of the original QTC, providing an independent test of the $K = 25/8$ BKT criterion.

\begin{figure}[t]
\centering
\includegraphics[width=0.97\columnwidth]{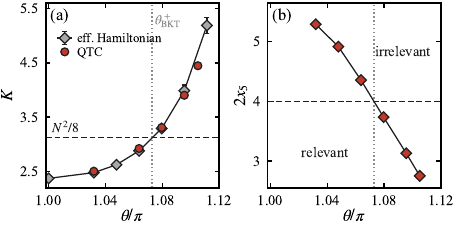}
\caption{Numerical evidence for the clock-anisotropy BKT transition in the $N = 5$ QTC near $\theta = \pi$. (a) Luttinger parameter $K_5$ obtained from the effective Hamiltonian $H_{\rm eff}^{(4,{\rm c})} (\delta)$ and the original QTC. Only $\theta > \pi$ is shown, owing to the exact $\delta \leftrightarrow -\delta$ symmetry of $H_{\rm eff}^{(4,{\rm c})} (\delta)$. The dashed line marks the clock-anisotropy threshold $K_c = 25/8$. (b) Scaling exponent $2x_5$ extracted from the connected $O_5$ correlation. The dashed line marks $2x_5 = 4$.}
\label{fig:BKT_O5}
\end{figure}

We now turn to the numerical results. Figure~\ref{fig:BKT_O5}(a) compares the Luttinger parameter $K_5$ obtained from $H_{\rm eff}^{(4,{\rm c})}(\delta)$ with that extracted from the original QTC. The effective Hamiltonian $H_{\rm eff}^{(4,{\rm c})}(\delta)$ has an exact $\delta \leftrightarrow -\delta$ symmetry [Eq.~\eqref{Heff4c_sym}], which enforces $K_5 (\delta) = K_5 (-\delta)$. Accordingly, we show only the $\theta > \pi$ side. The two estimates for $K_5$ agree closely and both reach the clock-anisotropy threshold $K_c = 25/8$ at $\theta_{\rm BKT}^{+} \simeq (1 + 0.0725)\pi$. To further test this criterion directly in the original QTC, Fig.~\ref{fig:BKT_O5}(b) shows the scaling exponent $2x_5$ extracted from the connected $O_5$ correlation. The exponent crosses the marginal value $2x_5 = 4$ at $\theta_{\rm BKT}^{+} \simeq (1 + 0.0728)\pi$, in close agreement with $\theta_{\rm BKT}^+ \simeq (1 + 0.0725)\pi$ inferred from $H_{\rm eff}^{(4,{\rm c})}(\delta)$. These results provide direct microscopic support for the BKT transitions driven by clock anisotropy at both boundaries of the dLL$_{\rm III}$ phase.

\SMsection{Web of dualities}{app:web_of_dualities}
We establish the duality web in two steps, mapping the dTCS first to the dTIM and subsequently to the QTC. The duality operators defined below use periodic site indices and yield bulk operator relations that hold for arbitrary $N$. On a finite ring, however, these operators intertwine only specific symmetry sectors, while the remaining sectors require appropriate boundary twists.

\SMsubsection{From dTCS to dTIM}{app:dTCS-to-dTIM}
The type-I dTCS and dTIM Hamiltonians are given explicitly by
\begin{align}
    H_{\rm dTCS,I}&=-\cos\theta\sum_j(S_j+S_j^\dagger) -\sin\theta\sum_j(X_j+X_j^\dagger), \nonumber \\
    S_j&=(Z_{j-1}Z_j^\dagger)X_j(Z_j^\dagger Z_{j+1}), \label{dTCS}\\
    H_{\rm dTIM}&=-\cos\theta\sum_j(D_j+D_j^\dagger) -\sin\theta\sum_j(X_j+X_j^\dagger),\nonumber\\
    D_j&=Z_{j-1}^\dagger Z_j^2Z_{j+1}^\dagger\,, \label{dTIM}
\end{align}
respectively. Both models possess monopole ($Q=\prod_jX_j$) and dipole ($D=\prod_jX_j^j$) symmetries. Under periodic boundary conditions, we take the system size $L$ to be a multiple of $N$ to preserve the full $\mathbb Z_N\times\mathbb Z_N$ charge–dipole symmetry.
The two models are related by the noninvertible transformation~\cite{cao24generating,han25noninvertible}
\begin{align}
     \mathcal M_{\mathrm I} =U_{\mathrm{DDW}}^\dagger K_{\mathrm{dKW}}U_{\mathrm{DDW}}^\dagger,
     \label{web:mi}
\end{align}
where the decorated-domain-wall operator~\cite{han24topological} acts in the $Z$ basis as
\begin{align}
    U_{\mathrm{DDW}}|\mathbf g\rangle =\omega^{\sum_j g_j(g_{j+1}-g_j)}|\mathbf g\rangle, \qquad |\mathbf g\rangle=\bigotimes_j|g_j\rangle,
     \label{web:uddw}
\end{align}
and the dipolar Kramers--Wannier operator is defined by
\begin{align}
    K_{\mathrm{dKW}} =\sum_{\mathbf g,\mathbf g'} \omega^{\sum_j(g_{j+1}-2g_j+g_{j-1})g'_j} |\mathbf g'\rangle\langle\mathbf g|,
     \label{web:kdkw}
\end{align}
which obeys the intertwining relations
\begin{align}
    K_{\mathrm{dKW}}X_j=D_j^\dagger K_{\mathrm{dKW}}, \quad K_{\mathrm{dKW}}D_j=X_jK_{\mathrm{dKW}}.
    \label{web:dkwrules}
\end{align}
Under $\mathcal M_{\mathrm I}$, the $X_j$ and $X_j^\dagger$ transform as
\begin{align}
     \mathcal M_{\mathrm I}X_j=X_j\mathcal M_{\mathrm I}, \quad \mathcal M_{\mathrm I}X_j^\dagger=X_j^\dagger\mathcal M_{\mathrm I}.
     \label{web:xrules}
\end{align}
Consequently, the transverse-field term remains invariant,
\begin{align}
     \sum_j(X_j+X_j^\dagger) \xrightarrow{\mathcal M_{\mathrm I}} \sum_j(X_j+X_j^\dagger).
     \label{web:field-sum}
\end{align}
For the cluster stabilizers, we obtain
\begin{align}
     \mathcal M_{\mathrm I}S_j=D_j^\dagger\mathcal M_{\mathrm I}, \quad \mathcal M_{\mathrm I}S_j^\dagger=D_j\mathcal M_{\mathrm I}.
     \label{web:mirules}
\end{align}
Summing over all sites maps the cluster interaction to the dipolar interaction,
\begin{align}
     \sum_j(S_j+S_j^\dagger) \xrightarrow{\mathcal M_{\mathrm I}} \sum_j(D_j+D_j^\dagger).
     \label{web:target-sums}
\end{align}

The type-II dTCS Hamiltonian shares the same transverse-field term, with cluster stabilizers given by
\begin{equation}
     \begin{aligned}
          S_{2j}&=Z_{2j-1}X_{2j}Z_{2j+1}^{-2}Z_{2j+3},\\
          S_{2j+1}&=Z_{2j-2}Z_{2j}^{-2}X_{2j+1}Z_{2j+2}.
     \end{aligned}
     \label{type-II-stabilizers}
\end{equation}
For each sublattice $a=0,1$, the independent symmetry generators are $Q^{(a)}=\prod_jX_{2j+a}$ and $D^{(a)}=\prod_jX_{2j+a}^{\,j}$. Preserving the full periodic dipole symmetry on each sublattice requires $L/2=0\pmod N$.
For an even chain length $L$, we introduce the sublattice operators $Z_{a,j}=Z_{2j+a}$ and $X_{a,j}=X_{2j+a}$ ($a=0,1$). The two corresponding sublattice dTIM Hamiltonians take the form
\begin{equation}
     \begin{aligned}
          H_{\mathrm{dTIM}}^{(a)} &=-\cos\theta\sum_j(D_{a,j}+D_{a,j}^\dagger)\\
          &\quad-\sin\theta\sum_j(X_{a,j}+X_{a,j}^\dagger).
     \end{aligned}
     \label{web:parity}
\end{equation}
Here $D_{a,j}=Z_{a,j-1}^\dagger Z_{a,j}^{2}Z_{a,j+1}^\dagger$.
For the type-II model, the noninvertible transformation is given by~\cite{han25noninvertible}
\begin{equation}
     \begin{aligned}
          \mathcal M_{\mathrm{II}}&=U_{\mathrm{II}}K_{\mathrm{II}}U_{\mathrm{II}}^\dagger,\\
          K_{\mathrm{II}}&=P K_{\mathrm{dKW}}^{(1)}(K_{\mathrm{dKW}}^{(0)})^\dagger,\\
          P&=\prod_j\operatorname{SWAP}_{2j,2j+1},
     \end{aligned}
     \label{web:mii}
\end{equation}
where the type-II decorated-domain-wall operator acts as
\begin{align}
    U_{\mathrm{II}}|\mathbf g\rangle =\omega^{\sum_j g_{2j}(g_{2j-1}-2g_{2j+1}+g_{2j+3})}|\mathbf g\rangle .
    \label{web:uii}
\end{align}
Here $K_{\mathrm{dKW}}^{(a)}$ denotes the operator in Eq.~\eqref{web:kdkw} acting on sublattice $a$, and $P$ exchanges the two sites within each unit cell. From Eq.~\eqref{web:dkwrules}, it follows that
\begin{equation}
     \begin{aligned}
          K_{\mathrm{II}}X_{0,j}&=D_{1,j}K_{\mathrm{II}},& K_{\mathrm{II}}X_{1,j}&=D_{0,j}^\dagger K_{\mathrm{II}},\\
          K_{\mathrm{II}}D_{0,j}&=X_{1,j}^\dagger K_{\mathrm{II}},& K_{\mathrm{II}}D_{1,j}&=X_{0,j}K_{\mathrm{II}}.
     \end{aligned}
     \label{web:kiirules}
\end{equation}
Under $\mathcal M_{\mathrm{II}}$, the transverse-field operators satisfy
\begin{equation}
     \begin{aligned}
          \mathcal M_{\mathrm{II}}X_{0,j}&=X_{0,j}\mathcal M_{\mathrm{II}}, &\quad \mathcal M_{\mathrm{II}}X_{0,j}^\dagger&=X_{0,j}^\dagger\mathcal M_{\mathrm{II}},\\
          \mathcal M_{\mathrm{II}}X_{1,j}&=X_{1,j}^\dagger\mathcal M_{\mathrm{II}}, &\quad \mathcal M_{\mathrm{II}}X_{1,j}^\dagger&=X_{1,j}\mathcal M_{\mathrm{II}}.
         \end{aligned}
     \label{web:iixrules}
\end{equation}
Although the odd-sublattice operators are mapped to their Hermitian conjugates, the total transverse-field term remains unchanged,
\begin{align}
     \sum_j(X_j+X_j^\dagger) \xrightarrow{\mathcal M_{\mathrm{II}}} \sum_{a,j}(X_{a,j}+X_{a,j}^\dagger).
     \label{web:ii-field-sum}
\end{align}
For the cluster stabilizers, we similarly find
\begin{equation}
     \begin{aligned}
          \mathcal M_{\mathrm{II}}S_{2j}&=D_{1,j}\mathcal M_{\mathrm{II}}, &\quad \mathcal M_{\mathrm{II}}S_{2j}^\dagger&=D_{1,j}^\dagger\mathcal M_{\mathrm{II}},\\
          \mathcal M_{\mathrm{II}}S_{2j+1}&=D_{0,j}^\dagger\mathcal M_{\mathrm{II}}, &\quad \mathcal M_{\mathrm{II}}S_{2j+1}^\dagger&=D_{0,j}\mathcal M_{\mathrm{II}}.
         \end{aligned}
     \label{web:miirules}
\end{equation}
Summing over both sublattices yields the two dipolar interactions,
\begin{align}
     \sum_j(S_j+S_j^\dagger) \xrightarrow{\mathcal M_{\mathrm{II}}} \sum_{a,j}(D_{a,j}+D_{a,j}^\dagger).
     \label{web:ii-sums}
\end{align}

\SMsubsection{From dTIM to QTC}{app:dTIM-to-QTC}
The dTIM is mapped to the QTC by the noninvertible Kramers--Wannier (KW) operator~\cite{kramers41statistics,cao24generating,han25noninvertible},
\begin{align}
    K_{\mathrm{KW}} =\sum_{\mathbf g,\mathbf g'} \omega^{\sum_j(g_{j-1}-g_j)g'_{j-1}} |\mathbf g'\rangle\langle\mathbf g| .
    \label{web:kkw}
\end{align}
Its action on the transverse-field operators is given by
\begin{equation}
     \begin{aligned}
          K_{\mathrm{KW}}X_j&=(Z_{j-1}^\dagger Z_j)K_{\mathrm{KW}},\\
          K_{\mathrm{KW}}X_j^\dagger&=(Z_{j-1}Z_j^\dagger)K_{\mathrm{KW}}.
     \end{aligned}
     \label{web:kwrules}
\end{equation}
which maps the transverse-field term to a nearest-neighbor $Z$ interaction,
\begin{align}
     \sum_j(X_j+X_j^\dagger) \xrightarrow{K_{\mathrm{KW}}} \sum_j(Z_{j-1}^\dagger Z_j+Z_{j-1}Z_j^\dagger).
     \label{web:kw-field-sum}
\end{align}
For the dipolar operators, we obtain
\begin{equation}
     \begin{aligned}
          K_{\mathrm{KW}}D_j&=X_{j-1}X_j^\dagger K_{\mathrm{KW}},\\
          K_{\mathrm{KW}}D_j^\dagger&=X_{j-1}^\dagger X_jK_{\mathrm{KW}}.
         \end{aligned}
     \label{web:kw-dipole}
\end{equation}
The dipolar interaction is thus mapped to a nearest-neighbor $X$ interaction,
\begin{align}
     \sum_j(D_j+D_j^\dagger) \xrightarrow{K_{\mathrm{KW}}} \sum_j(X_{j-1}X_j^\dagger+X_{j-1}^\dagger X_j).
     \label{web:kw-dipole-sum}
\end{align}
Combining these two relations yields the QTC Hamiltonian at the same parameter $\theta$,
\begin{align}
    K_{\mathrm{KW}}H_{\mathrm{dTIM}}(\theta) =H_{\mathrm{QTC}}(\theta)K_{\mathrm{KW}}.
    \label{web:hqtc}
\end{align}
In the type-II case, applying the KW transformation to each sublattice maps the model to two decoupled QTC chains,
\begin{equation}
     \begin{aligned}
          H_{\rm dTCS,II}&\xrightarrow{\mathcal M_{\rm II}} H_{\rm dTIM}^{(0)}+H_{\rm dTIM}^{(1)}\\
          &\xrightarrow{K_{\rm KW}^{(0)}\otimes K_{\rm KW}^{(1)}} H_{\rm QTC}^{(0)}+H_{\rm QTC}^{(1)}.
     \end{aligned}
     \label{app:full-two-copy-web}
\end{equation}
On a periodic QTC chain, the symmetry generators satisfy $U_ZU_X=\omega^L U_XU_Z$, and therefore commute whenever $L$ is a multiple of $N$.

\SMsubsection{Mapping of correlators}{app:cluster-correlators}
\label{app:torus-correlators}
The duality transformations established above also relate the order parameters of the three models. Specifically, the dTCS cluster string maps to a dipolar correlator in the dTIM and subsequently to a two-point $X$ correlator in the QTC, while the dTCS paramagnetic string maps to the $X$ string in the dTIM and then to a two-point $Z$ correlator in the QTC.

For the type-I dTCS, we define the cluster and paramagnetic string correlators as
\begin{equation}
     \begin{aligned}
          C_{{\rm dTCS}}^{\mathrm{cl}}(i,r) &=\left\langle\prod_{j=i}^{i+r-1}S_j \right\rangle_{\mathrm{dTCS}_{\mathrm I}},\\
          C_{{\rm dTCS}}^{\mathrm{pm}}(i,r) &=\left\langle\prod_{j=i}^{i+r-1}X_j^\dagger \right\rangle_{\mathrm{dTCS}_{\mathrm I}}.
     \end{aligned}
     \label{corr:istrings}
\end{equation}
Multiplying the stabilizers along the interval cancels the interior $Z$ factors, leaving endpoint operators attached to the $X$ string,
\begin{align}
     \prod_{j=i}^{i+r-1}S_j =(Z_{i-1}Z_i^\dagger) \left(\prod_{j=i}^{i+r-1}X_j\right) (Z_{i+r-1}^\dagger Z_{i+r}).
     \label{corr:endpoints}
\end{align}
To express the corresponding dTIM correlators, we introduce the bond variables
\begin{align}
     \widetilde Z_j=Z_{j-1}^\dagger Z_j, \qquad D_j=\widetilde Z_j\widetilde Z_{j+1}^\dagger .
     \label{corr:dualvariables}
\end{align}
in terms of which the dipolar and string correlators read
\begin{equation}
     \begin{aligned}
          C^D_{\mathrm{dTIM}}(i,r) &=\langle\widetilde Z_i^\dagger\widetilde Z_{i+r} \rangle_{\mathrm{dTIM}},\\
          C^S_{\mathrm{dTIM}}(i,r) &=\left\langle\prod_{j=i}^{i+r-1}X_j^\dagger \right\rangle_{\mathrm{dTIM}}.
     \end{aligned}
     \label{corr:dtim}
\end{equation}
Applying the local relations in Eqs.~\eqref{web:xrules} and \eqref{web:mirules} factor by factor gives
\begin{equation}
     \begin{aligned}
          \mathcal M_{\mathrm I}\left(\prod_{j=i}^{i+r-1}S_j\right)&=\widetilde Z_i^\dagger\widetilde Z_{i+r}\mathcal M_{\mathrm I},\\
          \mathcal M_{\mathrm I}\left(\prod_{j=i}^{i+r-1}X_j^\dagger\right)&=\left(\prod_{j=i}^{i+r-1}X_j^\dagger\right)\mathcal M_{\mathrm I}.
     \end{aligned}
     \label{corr:iproducts}
\end{equation}
Taking expectation values in the corresponding ground states then gives the following correlator relations:
\begin{equation}
     \begin{aligned}
          C_{\rm dTCS}^{\mathrm{cl}}&=C^D_{\mathrm{dTIM}},\\
          C_{\rm dTCS}^{\mathrm{pm}}&=C^S_{\mathrm{dTIM}},
     \end{aligned}
     \label{corr:itodtim}
\end{equation}
where all correlators share the same arguments $(i,r)$. These relations are understood in the thermodynamic limit, where boundary effects vanish.

For the type-II models, we analogously define the sublattice dTIM correlators as
\begin{equation}
     \begin{aligned}
          \widetilde Z_{a,j}&=Z_{a,j-1}^\dagger Z_{a,j},\\
          C_D^{(a)}(j,R)&=\langle\widetilde Z_{a,j}^\dagger \widetilde Z_{a,j+R}\rangle_{\mathrm{dTIM}},\\
          C_S^{(a)}(j,R)&=\left\langle\prod_{\ell=j}^{j+R-1} X_{a,\ell}^\dagger\right\rangle_{\mathrm{dTIM}}.
     \end{aligned}
     \label{corr:paritydtim}
\end{equation}
By Eqs.~\eqref{web:iixrules} and \eqref{web:miirules}, the corresponding type-II dTCS string correlators are given by
\begin{equation}
     \begin{aligned}
          \mathcal O_{\mathrm{II}}^{\mathrm{cl},0}(n,R) &=\left\langle\prod_{\ell=n}^{n+R-1}S_{2\ell+1}\right\rangle_{\mathrm{dTCS}_{\mathrm{II}}},\\
          \mathcal O_{\mathrm{II}}^{\mathrm{cl},1}(n,R) &=\left\langle\prod_{\ell=n}^{n+R-1}S_{2\ell}^\dagger\right\rangle_{\mathrm{dTCS}_{\mathrm{II}}},\\
          \mathcal O_{\mathrm{II}}^{\mathrm{pm},0}(n,R) &=\left\langle\prod_{\ell=n}^{n+R-1}X_{2\ell}^\dagger\right\rangle_{\mathrm{dTCS}_{\mathrm{II}}},\\
          \mathcal O_{\mathrm{II}}^{\mathrm{pm},1}(n,R) &=\left\langle\prod_{\ell=n}^{n+R-1}X_{2\ell+1}\right\rangle_{\mathrm{dTCS}_{\mathrm{II}}}.
     \end{aligned}
     \label{corr:iistrings}
\end{equation}
Applying $\mathcal M_{\mathrm{II}}$ to each factor in the product directly yields
\begin{equation}
     \begin{aligned}
          \mathcal O_{\mathrm{II}}^{\mathrm{cl},a}(n,R)&=C_D^{(a)}(n,R),\\
          \mathcal O_{\mathrm{II}}^{\mathrm{pm},a}(n,R)&=C_S^{(a)}(n,R).
     \end{aligned}
     \label{corr:iitodtim}
\end{equation}

Under the KW transformation, the dTIM correlator operators map to two-point endpoint operators in the QTC,
\begin{equation}
     \begin{aligned}
          K_{\mathrm{KW}}(\widetilde Z_i^\dagger\widetilde Z_{i+r})&=(X_{i-1}^\dagger X_{i+r-1})K_{\mathrm{KW}},\\
          K_{\mathrm{KW}}\left(\prod_{j=i}^{i+r-1}X_j^\dagger\right)&=(Z_{i-1}Z_{i+r-1}^\dagger)K_{\mathrm{KW}}.
     \end{aligned}
     \label{corr:kwproducts}
\end{equation}
We define the resulting two-point QTC correlators as
\begin{equation}
     \begin{aligned}
          C^X_{\mathrm{QTC}}(i,r) &=\langle X_{i-1}^\dagger X_{i+r-1}\rangle_{\mathrm{QTC}},\\
          C^Z_{\mathrm{QTC}}(i,r) &=\langle Z_{i-1}Z_{i+r-1}^\dagger\rangle_{\mathrm{QTC}}.
     \end{aligned}
\end{equation}
Combining both duality transformations then gives the complete correlator correspondence for the type-I chain,
\begin{equation}
     \begin{alignedat}{2}
          C_{\rm dTCS}^{\mathrm{cl}}(i,r) &=C^D_{\mathrm{dTIM}}(i,r)&&=C^X_{\mathrm{QTC}}(i,r),\\
          C_{\rm dTCS}^{\mathrm{pm}}(i,r) &=C^S_{\mathrm{dTIM}}(i,r)&&=C^Z_{\mathrm{QTC}}(i,r).
         \end{alignedat}
     \label{complete-correlator-dictionary}
\end{equation}
Similarly, for the type-II models, defining the sublattice QTC correlators as
\begin{equation}
     \begin{aligned}
          C_X^{\mathrm{QTC},(a)}(n,R) &=\langle X_{a,n-1}^\dagger X_{a,n+R-1}\rangle_{\mathrm{QTC}},\\
          C_Z^{\mathrm{QTC},(a)}(n,R) &=\langle Z_{a,n-1}Z_{a,n+R-1}^\dagger\rangle_{\mathrm{QTC}}.
         \end{aligned}
     \label{corr:parityqtc}
\end{equation}
yields the sublattice correlator correspondence
\begin{equation}
     \begin{alignedat}{2}
          \mathcal O_{\mathrm{II}}^{\mathrm{cl},a}(n,R) &=C_D^{(a)}(n,R)&&=C_X^{\mathrm{QTC},(a)}(n,R),\\
          \mathcal O_{\mathrm{II}}^{\mathrm{pm},a}(n,R) &=C_S^{(a)}(n,R)&&=C_Z^{\mathrm{QTC},(a)}(n,R).
     \end{alignedat}
     \label{corr:iidictionary}
\end{equation}

\SMsection{Additional numerical evidence}{app:add_numerical_evidence}

\SMsubsection{Entanglement entropy within SO and C phases}{app:ent_entropy_SO_C}

\begin{figure}[t]
\centering
\includegraphics[width=0.95\columnwidth]{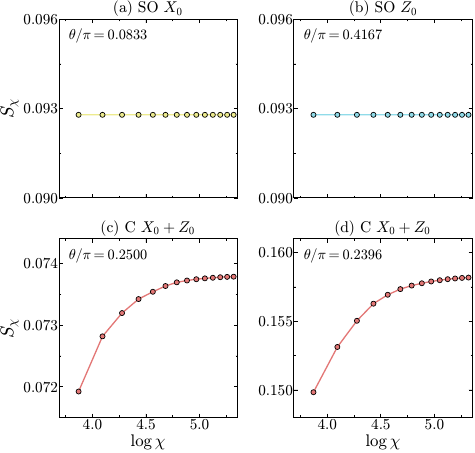}
\caption{Entanglement entropy $S_\chi$ as a function of logarithm of the MPS bond dimension $\chi$ for the $N = 8$ QTC: (a) $\theta = \pi/12$ in the $X_0$ SO phase, (b) $\theta = 5\pi/12$ in the $Z_0$ SO phase, and (c) $\theta = \pi/4$ and (d) $\theta = 23\pi/96$ in the $X_0+Z_0$ C phase. In all cases, $S_\chi$ converges to a finite value as $\chi$ increases, consistent with area-law scaling in the SO and C phases.}
\label{fig:SM_EE_SO_C}
\end{figure}

We first examine the entanglement entropy $S_\chi$ as a function of the MPS bond dimension $\chi$ for representative points within the SO and C phases. As shown in Fig.~\ref{fig:SM_EE_SO_C}, the entanglement entropy exhibits a clear bending and eventually saturates to a finite value as $\log \chi$ increases, consistent with area-law scaling. This saturation indicates that the correlation length approaches a finite value in the large-$\chi$ limit, supporting the gapped nature of both the SO and C phases.

\SMsubsection{Detailed numerical data for dipolar Luttinger liquid}{app:numerics_dLL}

\begin{figure}[t]
\centering
\includegraphics[width=\columnwidth]{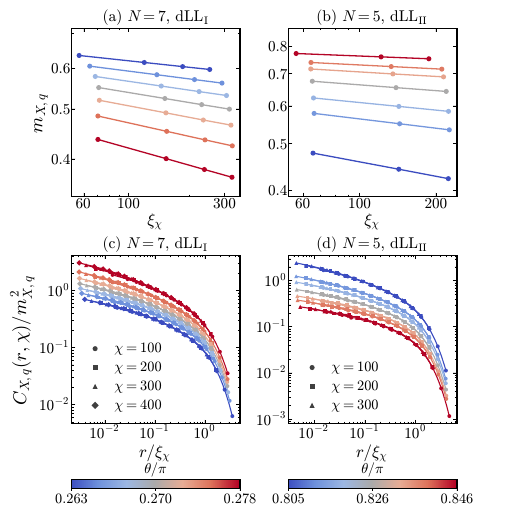}
\caption{Finite-entanglement scaling in dLL$_{\mathrm I}$ and dLL$_{\mathrm{II}}$. Left and right columns show $N = 7$, dLL$_{\mathrm I}$, and $N = 5$, dLL$_{\mathrm{II}}$, respectively. (a,b) Finite-$\chi$ pseudo-order parameter $m_{X,q}$ versus the correlation length $\xi_\chi$, with $q = 0$ for dLL$_{\mathrm I}$ and $q = \pi$ for dLL$_{\mathrm{II}}$. (c,d) $C_{X,q}(r,\chi) / m_{X,q}(\chi)^2$ as a function of $r / \xi_\chi$, with $C_{X,q}$ defined in Eq.~\eqref{CXq}. Colors denote $\theta / \pi$ within each column, and symbols in (c,d) identify $\chi$. The parameter-free collapse across bond dimensions at fixed $\theta$ supports finite-entanglement scaling.}
\label{fig:SM_dLL_FSS}
\end{figure}

We now provide further numerical evidence for the dLL$_{\rm I}$ and dLL$_{\rm II}$ phases. In both phases, the apparent $X_q$ order observed at finite bond dimension $\chi$ is expected to vanish in the thermodynamic limit. To demonstrate this, Figs.~\ref{fig:SM_dLL_FSS}(a) and (b) show the corresponding order parameters $m_{X,q}$ as functions of the correlation length $\xi_\chi$ for dLL$_{\rm I}$ ($N = 7$, $q = 0$) and dLL$_{\rm II}$ ($N = 5$, $q = \pi$), respectively. The data exhibit clear power-law behavior of the form
\begin{align}
    m_{X,q} (\chi) \sim \xi_\chi^{-\Delta_X},
\end{align}
with a scaling dimension $\Delta_X$. Thus, the apparent $X_q$ order decreases systematically with increasing $\chi$ and vanishes as $\xi_\chi \to \infty$, confirming that it is a finite-entanglement pseudo-order rather than genuine long-range order. The scaling dimension is related to the algebraic decay exponent $\eta_X$ of the $XX$ correlations through $2\Delta_X = \eta_X = 1/(2K)$ (see Fig.~\ref{fig:dLL_evidence} of the main text), providing a direct connection to the Luttinger-liquid description.

As an independent test of finite-entanglement scaling, we consider the connected correlation of the modulated operator $\widetilde{X}_j = e^{iqj} X_j$:
\begin{align} \label{CXq}
    C_{X,q} (r,\chi) = \langle \widetilde{X}_0 \widetilde{X}_r^\dagger \rangle_\chi - | m_{X,q} (\chi) |^2,
\end{align}
where $\langle \cdot \rangle_\chi$ denotes the expectation value in the MPS with bond dimension $\chi$. At fixed $\theta$, finite-entanglement scaling predicts
\begin{align}
    C_{X,q} (r,\chi) = \xi_\chi^{-2\Delta_X} \mathcal{F}_{X,q,\theta} (r / \xi_\chi),
\end{align}
where $\mathcal{F}_{X,q,\theta}$ is a scaling function. Together with $m_{X,q} (\chi) \sim \xi_\chi^{-\Delta_X}$, this implies
\begin{align}
    \frac{C_{X,q} (r,\chi)}{m_{X,q} (\chi)^2} = \widetilde{\mathcal{F}}_{X,q,\theta} (r / \xi_\chi),
\end{align}
so that $C_{X,q} (r,\chi) / m_{X,q} (\chi)^2$ becomes a function of the scaled distance $r / \xi_\chi$ alone. Figures~\ref{fig:SM_dLL_FSS}(c) and
(d) show this parameter-free collapse for dLL$_{\rm I}$ ($N = 7$) and dLL$_{\rm II}$ ($N = 5$), respectively. For each $\theta$, data obtained at different bond dimensions collapse onto a common curve without any adjustable parameters. This collapse provides a nontrivial consistency check of the finite-entanglement scaling of both the order parameter and the connected correlation, further supporting the identification of the two phases as dipolar Luttinger liquids.

\end{document}